\documentclass[referee]{raa}            % referee version: for submission

\usepackage{graphicx,times}             %for PS/EPS graphics inclusion, new

\begin{document}

\title{$C$-field cosmological models: revisited}

 \volnopage{ {\bf 2012} Vol.\ {\bf X} No. {\bf XX}, 000--000}
   \setcounter{page}{1}

 \author{A.K. Yadav\inst{1} \and A.T. Ali\inst{2} \and Saibal Ray\inst{3} \and F. Rahaman\inst{4} \and I.H. Sardar\inst{4} }

\institute{Department of Physics, United College of Engineering
and Research, Greater Noida 201306, India; {\it
abanilyadav@yahoo.co.in}\\ \and Department of Mathematics, King
Abdul Aziz University, PO Box 80203, Jeddah 21589, Saudi Arabia \&
Department of Mathematics, Faculty of Science, Al-Azhar
University, Nasr city, 11884, Cairo, Egypt; {\it
atali71@yahoo.com}\\ \and Department of Physics, Government
College of Engineering and Ceramic Technology, Kolkata 700010,
West Bengal, India; {\it saibal@associates.iucaa.in}\\ \and
Department of Mathematics, Jadavpur University, Kolkata 700032,
West Bengal,India; {\it rahaman@associates.iucaa.in}\\ \and
Department of Mathematics, Jadavpur University, Kolkata 700032,
West Bengal,India; {\it iftikar.spm@gmail.com} \\ }

\date{Received~~2016 month day; accepted~~2016~~month day}

\abstract{We investigate plane symmetric spacetime filled with
perfect fluid in the $C$-field cosmology of Hoyle and Narlikar. A
new class of exact solutions have been obtained by considering the
creation field $C$ as a function of time only. To get the
deterministic solution, it has been assumed that the rate of
creation of matter-energy density is proportional to the strength
of the existing $C$-field energy density. Several physical aspects
and geometrical properties of the models are discussed in detail,
especially it is shown that some of our solutions of $C$-field
cosmology are free from singularity in contrast to the Big Bang
cosmology. A comparative study has been carried out between two
models, one singular and the other nonsingular, by contrasting the
behaviour of the physical parameters and noted that the model in a
unique way represents both the features of the accelerating as
well as decelerating Universe depending on the parameters and thus
seems provides glimpses of the oscillating or cyclic model of the
Universe without invoking any other agent or theory in allowing
cyclicity.
\keywords{cosmology: miscellaneous; cosmology: theory; cosmology: early universe}
}

\authorrunning{A.K. Yadav et al.}
\titlerunning{$C$-field cosmology}

\maketitle

\section{Introduction}

It is generally accepted that spatial anisotropy and the lack of
homogeneity would have important consequences in the very early
universe. Therefore the study of creation field cosmological model
that relax the FRW assumptions is well motivated and thus argued
not only as a viable alternative to the standard big-bang model
but also theoretically superior to that model (Narlikar \& Padmanabhan~\cite{Narlikar1985}).
As an additional support for this superiority Narlikar and
Rana~(\cite{Narlikar1983}) earlier showed that the theoretical curve
of relic radiation in the $G$-varying Hoyle-Narlikar cosmology
provides an acceptable fit to the observations at long as well as
short wavelengths. A similar problem was also studied by Narlikar
et al.~(\cite{Narlikar2003}) to calculate the expected angular power
spectrum of the temperature fluctuations in the microwave
background radiation generated in the quasi steady state cosmology
and were able to obtain a satisfactory fit to the observational
band power estimates of the CMBR temperature fluctuation spectrum.
An exhaustive review on the steady state cosmology and $C$-field
may be helpful in this research arena (Hoyle \& Narlikar~\cite{Hoyle1995}).

However, the alternative theories have been proposed from time to
time - the most well known being the steady state theory of
cosmology proposed by Bondi and Gold~(\cite{Bondi1948}). In this
approach the universe does not have any singular beginning nor an
end on the cosmic time scale. It has been postulated that the
statistical properties of the large scale features of the
universes do not change.

Narlikar and Padmanabhan~(\cite{Narlikar1985}) earlier found out a
solution of Einstein's equations which admits radiation and a
negative-energy based massless scalar creation field as a source.
They have shown that the cosmological model connected to this
solution satisfies all the observational tests. The model obtained
by them was very important specifically being free from
singularity and it could provide a natural explanation for the
flatness problem. Motivated by this fundamental work, in the
present work we have studied the Hoyle-Narlikar $C$-field
cosmology in plane symmetric space-time. We have assumed that
$C(x,t)=C(t)$ i.e., the creation field $C$ is a function of time
only. We have extended the method used by Narlikar and
Padmanabhan~(\cite{Narlikar1985}) to the plane symmetric model.

In this regard we note that cosmological model exhibiting plane
symmetry have attracted much attention to several scientists. It
was Taub~(\cite{Taub1951,Taub1956}) who first discussed about
plane symmetric perfect fluid distribution in which the flow was
taken to be isentropic in general relativity. Later on, as a
particular case of the plane symmetric models for cosmology, the
Bianchi type space-time has been extensively studied by Heckmann
and Schucking~(\cite{Heckmann1962}), Thorne~(\cite{Thorne1967}),
Jacobs~(\cite{Jacobs1968}), Singh and Singh~(\cite{Singh1992}).

More elaborately, in connection to plane symmetric space-time
Smoot et al.~(\cite{Smoot1992}) argued that the earlier
predictions of the Friedman-Lema{\^i}tre-Robertson-Walker type
models do not always exactly  explain the observed results. Some
peculiar outcomes regarding the redshift from the extra galactic
objects continue to contradict the theoretical explanations given
from the FLRW model. It is further known that symmetry plays an
important role to understand the structure of the universe, as
such distance measurements are usually thought to probe the
background metric of the universe, but in reality the presence of
perturbations will lead to deviations from the result expected in
an exactly homogeneous and isotropic universe which suggests to
consider the cases where perturbations are plane symmetric~(Adamek
2014). Though most of the stars are believed to have spherical
symmetry, however, cylindrical and plane symmetries may be useful
to investigate the gravitational waves which have been detected
very recently. So in literature, many authors consider plane
symmetry, which is less restrictive than spherical symmetry and
provides an avenue to study inhomogeneities in early as well as
late universe in different physical contexts by Da Silva and
Wang~(1998), Anguige~(\cite{Anguige2000}), Nouri-Zonoz and
Tavanfar~(\cite{Nouri-Zonoz2001}), Pradhan et
al.~(\cite{Pradhan2003,Pradhan2007}), Yadav~(\cite{Yadav2011}).
All these have inspired  us to study the model of the universe
with plane symmetry.

However, as background of the creation field cosmology we would
like to present here some of the relevant works which will provide
thread of our investigation. In their paper on Mach's principle
and the creation of matter Hoyle and Narlikar~(\cite{Hoyle1963})
have used the experimental evidence that the local inertial frame
is the one with respect to which the distant parts of the universe
are non-rotating. They introduced a scalar `creation field' into
the theory of relativity to improve the situation and showed that
this explains the observed remarkable degree of homogeneity and
isotropy in the universe.

It has also been shown via a $C$-field that the steady-state
cosmology appears as an asymptotic case of the cosmological
solutions of Einstein's equations. The source equation has been
treated in terms of discrete particles instead of the macroscopic
case of a smooth fluid (Hoyle \& Narlikar~\cite{Hoyle1964a}). In this sequel of works
on Steady-State cosmology, Hoyle and Narlikar~(\cite{Hoyle1966}) also
shown that it is possible to interpret that (i) the expansion rate
of fluctuation from the steady-state situation follows the
Einstein-de Sitter relations, (ii) the natural scale set by the
new steady-state corresponds to the masses of clusters of galaxies
$10^{13}~M_{\odot}$ for the `observable universe', and (iii) it is
suggested that elliptical galaxies were formed early in the
development of a fluctuation. Some other works on $C$-field
cosmology are available in the
literature (Hoyle \& Narlikar~\cite{Hoyle1964b}; Hoyle \& Narlikar~\cite{Hoyle1964c};
Narlikar~\cite{Narlikar1973}) for further study.

Very recently a study has been carried out (Ghate \& Mhaske~\cite{Ghate2014a}) in
the Hoyle-Narlikar creation field theory of gravitation under
plane symmetric and LRS Bianchi type $V$ cosmological models.
The work is on varying gravitational constant $G$ for the
barotropic fluid distribution. The solution of the field equations
have been obtained by assuming that $G=Bm$, where $B$ is scale
factor and $m$ is a constant. Besides this one Ghate and his
collaborators (Ghate \& Salve~\cite{Ghate2014b,Ghate2014c,Ghate2014d}) have
published series of works under $C$-field cosmology with different
physical systems. Some other recent works on $C$-field cosmology
are also available in the
literature (Chatterjee \& Banerjee~\cite{Chatterjee2004}; Singh \& Chaubey~\cite{Singh2009};
Adhav et al.~\cite{Adhav2010,Adhav2011}; Bali \& Saraf~\cite{Bali2013}).

The plan of our study is as follows: In the Sec. 2 we have given
an overall view of the Creation field theory in cosmology whereas
in the Sec. 3 and Sec. 4 the basic mathematical details of the
model and exact solutions of the models respectively have been
provided. A special section has been added there after in Sec. 5
for the non-singular solution. We have discussed several physical
features of the models in the Sec. 6. In the last Sec. 7 we have
passed some concluding remarks based on comparative studies
between two models, one singular and the other nonsingular, by
contrasting the behaviour of different physical parameters.

\section{The Creation field theory}

Einstein's field equations are modified by introducing a mass less
scalar field called as creation field, viz. $C$-field
(Hoyle \& Narlikar~\cite{Hoyle1963,Hoyle1964a,Hoyle1964b,Hoyle1964c,Hoyle1966};
Narlikar~\cite{Narlikar1973}; Narlikar et al.~\cite{Narlikar2003}).
The proposed modified field equations have been provided in the
form
\begin{equation}  \label{u21}
R_{ij}-\frac{1}{2}\,g_{ij}\,R=-8\,\pi\,\Big(\,^mT_{ij}+\,^cT_{ij}\Big),
\end{equation}
where $^mT_{ij}$ is the matter tensor of the Einstein theory and
$^cT_{ij}$ is the matter tensor due to the $C$-field which is
given by
\begin{equation}  \label{u22}
^cT_{ij}=-f^2\Big(C_i\,C_j-\frac{1}{2}\,g_{ij}\,C^k\,C_k\Big),
\end{equation}
where $f^2$ is a coupling constant, $C_i=\frac{\partial
C}{\partial x^i}$ and $C$ is the creation field function. It is
not necessary to take small value of coupling constant $f$.
However, it is not large enough and hence one can assume the value
of $f$ in such a way that all the solutions have finite values.

Because of the negative value of $T^{00}$, the $C$-field has
negative energy density producing repulsive gravitational field
which causes the expansion of the universe. Hence, the energy
conservation equation reduces to
\begin{equation}  \label{u23}
^mT^{ij}_{;j}=-\,^cT^{ij}_{;j}=f^2\,C^i\,C_{;j}.
\end{equation}
Here the semicolon (;) denotes covariant differentiation, i.e. the
matter creation through non-zero left hand side is possible while
conserving the over all energy and momentum.

\section{The models: Mathematical basics}

The spatially homogeneous and anisotropic plane symmetric
space-time is described by the line element
\begin{equation}  \label{u31}
ds^2=dt^2-A^2\Big(dx^2+dy^2\Big)-B^2\,dz^2,
\end{equation}
where $A$ and $B$ are the cosmic scale factors and the functions
of the cosmic time $t$ only (non-static case).

The proper volume of the model (\ref{u31}) is given by
\begin{equation}  \label{u31-1}
V=\sqrt{-g}=A^2\,B.
\end{equation}
The matter tensor for perfect fluid is
\begin{equation}  \label{u32}
^mT^{ij}=diag(\rho,-p,-p,-p),
\end{equation}
where $\rho$ is the homogeneous mass density and $p$ is the
isotropic pressure. We have assumed here that the creation field
$C$ is function of time $t$ only i.e. $C(x,t) = C(t)$.

For the line element (\ref{u31}) the Einstein field equation
(\ref{u21}) can be written as
\begin{equation}  \label{u33}
\begin{array}{ll}
    8\,\pi\,\rho=4\,\pi\,\Omega+\frac{\dot{A}^2}{A^2}+\frac{2\,\dot{A}\,\dot{B}}{A\,B},
  \end{array}
\end{equation}

\begin{equation}  \label{u34}
\begin{array}{ll}
        8\,\pi\,p=4\,\pi\,\Omega-\frac{\dot{A}\dot{B}}{A\,B}-\frac{\ddot{B}}{B}-\frac{\ddot{A}}{A},
  \end{array}
\end{equation}

\begin{equation}  \label{u35}
\begin{array}{ll}
        \frac{\ddot{B}}{B}-\frac{\ddot{A}}{A}=
           \frac{\dot{A}}{A}\Big(\frac{\dot{A}}{A}-\frac{\dot{B}}{B}\Big),
  \end{array}
\end{equation}
where dot $( ^.)$ indicates the derivative with respect to $t$ and
$\Omega=f^2\,\dot{C}^2$. From (\ref{u31-1}), we can write
$B=\frac{V}{A^2}$. The equation (\ref{u35}) transforms to
\begin{equation}  \label{u35-1}
\begin{array}{ll}
        \frac{\ddot{V}}{3\,V}-\frac{\ddot{A}}{A}=
           \frac{\dot{A}}{A}\Big(\frac{\dot{V}}{V}-\frac{\dot{A}}{A}\Big),
  \end{array}
\end{equation}
The general solution of the above equation is
\begin{equation}  \label{u36}
\begin{array}{ll}
A(t)=a_1\,V^{1/3}(t)\,\exp\Big[a_2\,\int\frac{dt}{V(t)}\Big],
  \end{array}
\end{equation}
where $a_1$ and $a_2$ are constants of integration. Therefore, the
coefficient $B$, the homogeneous mass density $\rho$ and the
isotropic pressure become
\begin{equation}  \label{u37}
\begin{array}{ll}
B(t)=\frac{V^{1/3}(t)}{a_1^2}\,\exp\Big[-2\,a_2\,\int\frac{dt}{V(t)}\Big],
  \end{array}
\end{equation}

\begin{equation}  \label{u38}
\begin{array}{ll}
8\,\pi\,\rho(t)=4\,\pi\,\Omega(t)-\frac{3\,a_2^2}{V^2(t)}+\frac{\dot{V}^2(t)}{3\,V^2(t)},
  \end{array}
\end{equation}

\begin{equation}  \label{u39}
\begin{array}{ll}
8\,\pi\,p(t)=4\,\pi\,\Omega(t)-\frac{3\,a_2^2}{V^2(t)}+\frac{\dot{V}^2(t)}{3\,V^2(t)}-\frac{2\,\ddot{V}(t)}{3\,V(t)}.
  \end{array}
\end{equation}

In order to obtain a unique solution, one has to specify the rate
of creation of matter-energy (at the expense of the negative
energy of the $C$-field). Without loss of generality, we assume
that the rate of creation of matter energy density is proportional
to the strength of the existing $C$-field energy-density, i.e. the
rate of creation of matter energy density per unit proper-volume
is given by
\begin{equation}  \label{u310}
\begin{array}{ll}
\frac{d}{dV}\Big(\rho\,V\Big)+p=f^2\,\alpha^2\,\dot{C}^2,
  \end{array}
\end{equation}
where $\alpha$ is proportional constant.

The above equation can be written in the following form
\begin{equation}  \label{u310-1}
\begin{array}{ll}
V\,\dot{\rho}+\Big(p+\rho-\alpha^2\,\Omega\Big)\,\dot{V}=0.
  \end{array}
\end{equation}

Substituting Eqs. (\ref{u38}) and (\ref{u39}) in Eq. (\ref{u310-1}), we get
\begin{equation}  \label{u311}
\begin{array}{ll}
\frac{\dot{\Omega}}{\Omega}=2\,(\alpha^2-1)\,\frac{\dot{V}}{V}.
  \end{array}
\end{equation}

Integrating the above equation we have
\begin{equation}  \label{u312}
\begin{array}{ll}
\Omega(t)=\frac{\Omega_0}{4\,\pi}\,V^{2\,(\alpha^2-1)}\,,
  \end{array}
\end{equation}
where $\Omega_0$ is an arbitrary constant of integration. From
(\ref{u312}) in to (\ref{u38}) and (\ref{u39}) we have
\begin{equation}  \label{u38-1}
\begin{array}{ll}
8\,\pi\,\rho(t)=\Omega_0\,V^{2\,(\alpha^2-1)}(t)-\frac{3\,k_2^2}{V^2(t)}+\frac{\dot{V}^2(t)}{3\,V^2(t)},
  \end{array}
\end{equation}

\begin{equation}  \label{u39-1}
\begin{array}{ll}
8\,\pi\,p(t)=\Omega_0\,V^{2\,(\alpha^2-1)}(t)-\frac{3\,k_2^2}{V^2(t)}+\frac{\dot{V}^2(t)}{3\,V^2(t)}-\frac{2\,\ddot{V}(t)}{3\,V(t)}.
  \end{array}
\end{equation}

Now, we consider the equation of state of matter as
\begin{equation}  \label{u313}
\begin{array}{ll}
p=\gamma\,\rho,
  \end{array}
\end{equation}

Here $\gamma$ varies between the interval $0\leq\gamma\leq1$,
whereas $\gamma=0$ describes the dust universe, $\gamma=1/3$
presents the radiation universe, $1/3\leq\gamma\leq1$ describes
the hard universe and $\gamma=1$ corresponds to the stiff matter.

Substituting Eqs. (\ref{u313}) and (\ref{u312}) in Eq. (\ref{u310-1}), we get
\begin{equation}  \label{u310-2}
\begin{array}{ll}
V\,\dot{\rho}+\Big[(1+\gamma)\rho-\Omega_0\,\alpha^2\,V^{2(\alpha^2-1)}\Big]\dot{V}=0,
  \end{array}
\end{equation}
which yields
\begin{equation}  \label{u314}
\begin{array}{ll}
8\,\pi\,\rho(t)=\frac{2\,\Omega_0\,\alpha^2\,V^{2(\alpha^2-1)}}{2\,\alpha^2+\gamma-1}+\rho_0\,V^{-1-\gamma},
  \end{array}
\end{equation}
where $\rho_0$ is an arbitrary constant of integration.

Subtracting Eq. (\ref{u314}) from Eq. (\ref{u38-1}), we get
\begin{equation}  \label{u315}
\begin{array}{ll}
(2\,\alpha^2+\gamma-1)\Big[9\,a_2^2+3\,\rho_0\,V^{1-\gamma}-\dot{V}^2\Big]+3\,\Omega_0\,(1-\gamma)\,V^{2\,\alpha^2}=0.
  \end{array}
\end{equation}

The above equation can be written in the following form
\begin{equation}  \label{u315-1}
\begin{array}{ll}
\int
\frac{dV}{\sqrt{9\,a_2^2+k_0\,V^{2\,\alpha^2}+3\,\rho_0\,V^{1-\gamma}}}=t-t_0,
  \end{array}
\end{equation}
where $k_0=\frac{3\,\Omega_0\,(1-\gamma)}{2\,\alpha^2+\gamma-1}$
and $t_0$ is an arbitrary constant of integration.

\section{The models: A class of exact solutions}

To obtain the class of exact solution in terms of cosmic time $t$,
we consider the following cases and their respective plots. We
have used geometrical unit, i.e. $G=c=1$. The figures provide the
information of the nature variation of the physical parameters
with respect to time only. Usually the units are as follows:
energy density $\rightarrow gm/cm^3$, pressure $\rightarrow
dyne/cm^2$, creation field $C$ = density $\rightarrow gm/cm^3$,
volume $\rightarrow  cm^3$, time $\rightarrow$ Gyr.

\subsection{$\rho_0=0$}

\subsubsection{$a_2=0$} In this case, we can obtain the following
solution:

\begin{equation}  \label{u316}
\begin{array}{ll}
V(t)=\Big[k_1\,(1-\alpha^2)\,T\Big]^{\frac{1}{1-\alpha^2}},\,\,\,\,\,\,\,\,\,\,
\rho(t)=\frac{\alpha^2}{12\,\pi\,(1-\gamma)\,(1-\alpha^2)^2\,T^2},\\
\\
p(t)=\frac{\gamma\,\alpha^2}{12\,\pi\,(1-\gamma)\,(1-\alpha^2)^2\,T^2},\,\,\,\,\,
C(t)=C_0+\frac{1}{2\,f\,(1-\alpha^2)}\sqrt{\frac{2\,\alpha^2+\gamma-1}{3\,\pi\,(1-\gamma)}}\,\ln[T],\\
\\
A(t)=a_1\,\Big[(1-\alpha^2)\,T\Big]^{\frac{1}{3\,(1-\alpha^2)}},\,\,\,\,\,\,\,\,
B(t)=\frac{1}{a_1^2}\,\Big[k_1\,(1-\alpha^2)\,T\Big]^{\frac{1}{3\,(1-\alpha^2)}},
\end{array}
\end{equation}
where $C_0$ is an arbitrary constant, $k_0=k_1^2$ and $T=t-t_0$.

\begin{figure*}[thbp]
\begin{center}
\vspace{0.5cm}
\includegraphics[width=0.43\textwidth]{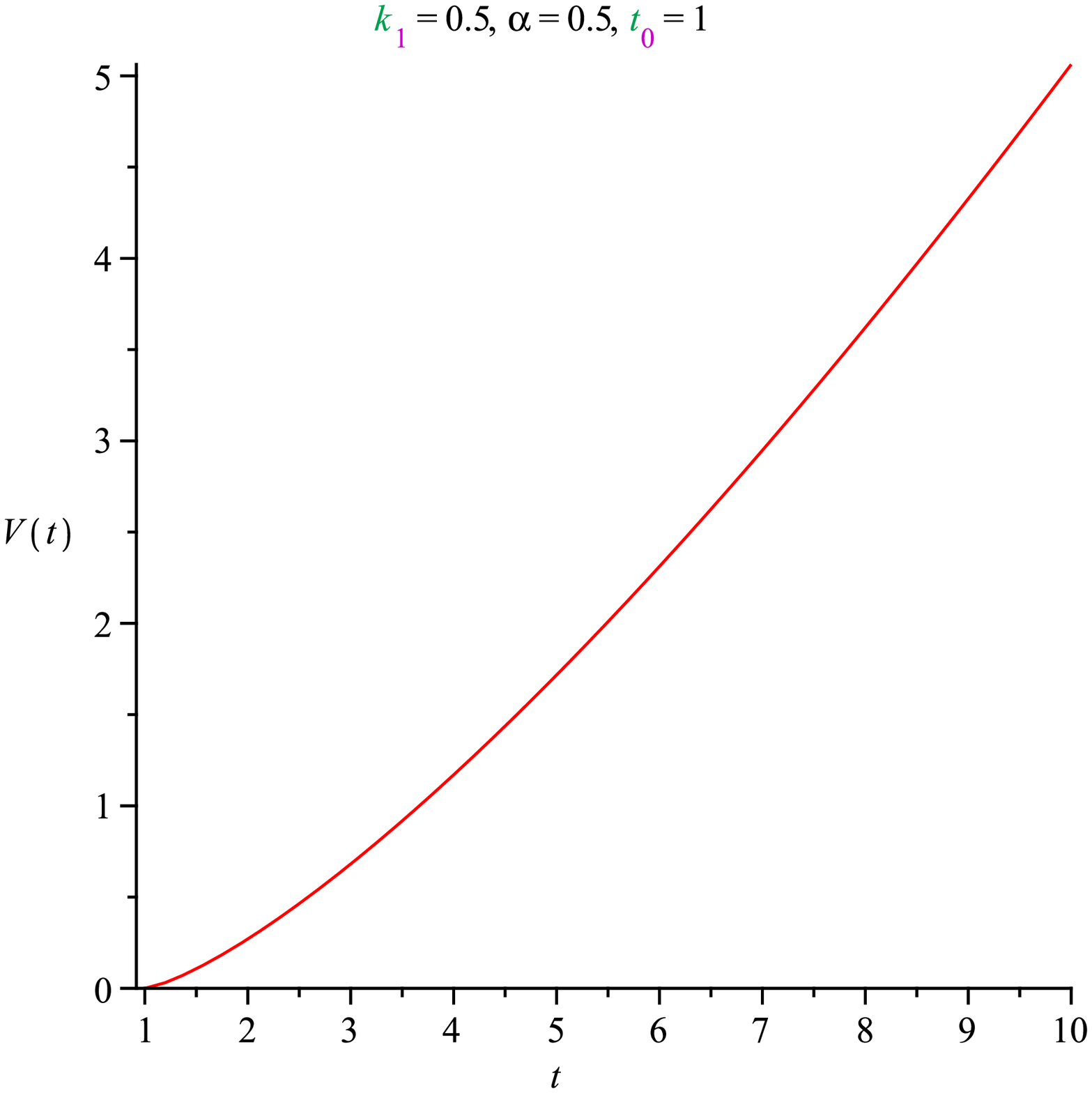}
\includegraphics[width=0.43\textwidth]{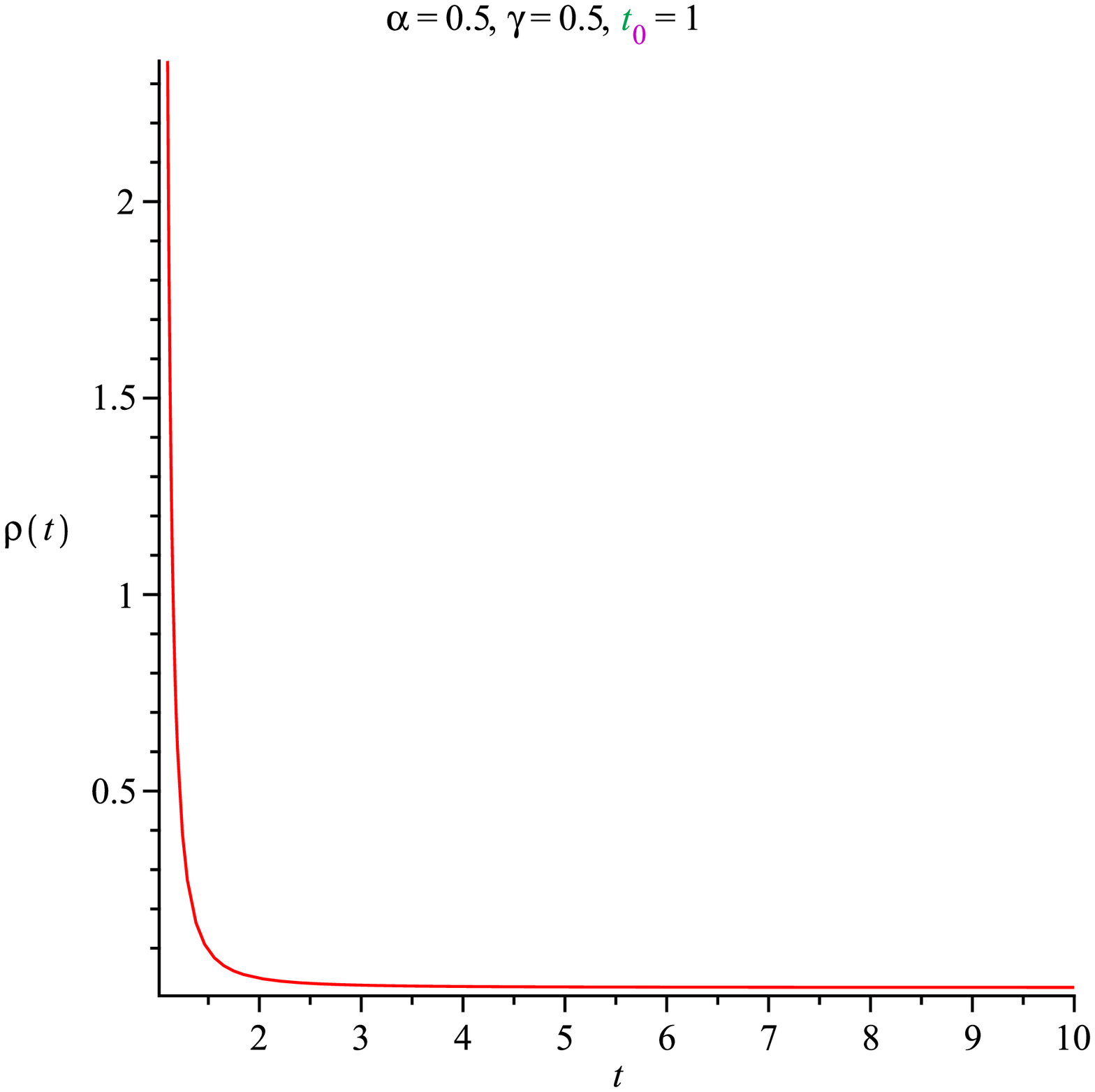}
\caption{Variation of volume (left panel) and density (right
panel) for Sub-case 4.1.1}
\end{center}
\end{figure*}

\subsubsection{$a_2 \neq 0$}

(i) For $\alpha=0$ case we can obtain the following solution:

\begin{equation}  \label{u317-1}
\begin{array}{ll}
V(t)=k_2\,T,\,\,\,\,\,\,\,\rho(t)=p(t)=0,\,\,\,\,\,\,\,C(t)=C_0+\frac{1}{2\,f\,k_2}\,\sqrt{\frac{9\,a_2^2-k_2^2}{3\,\pi}}\,\ln[T],\\
\\
A(t)=a_1\,k_2^{1/3}\,T^{\frac{1}{3}+\frac{a_2}{k_2}},\,\,\,\,\,\,\,\,\,\,\,\,\,\,\,
B(t)=\frac{k_2^{1/3}}{a_1^2}\,T^{\frac{1}{3}-\frac{2\,a_2}{k_2}},
\end{array}
\end{equation}
where $C_0$ is an arbitrary constant, $k_2^2=9\,a_2^2-3\,\Omega_0$
and $T=t-t_0$.

\begin{figure*}[thbp]
\begin{center}
\vspace{0.5cm}
\includegraphics[width=0.43\textwidth]{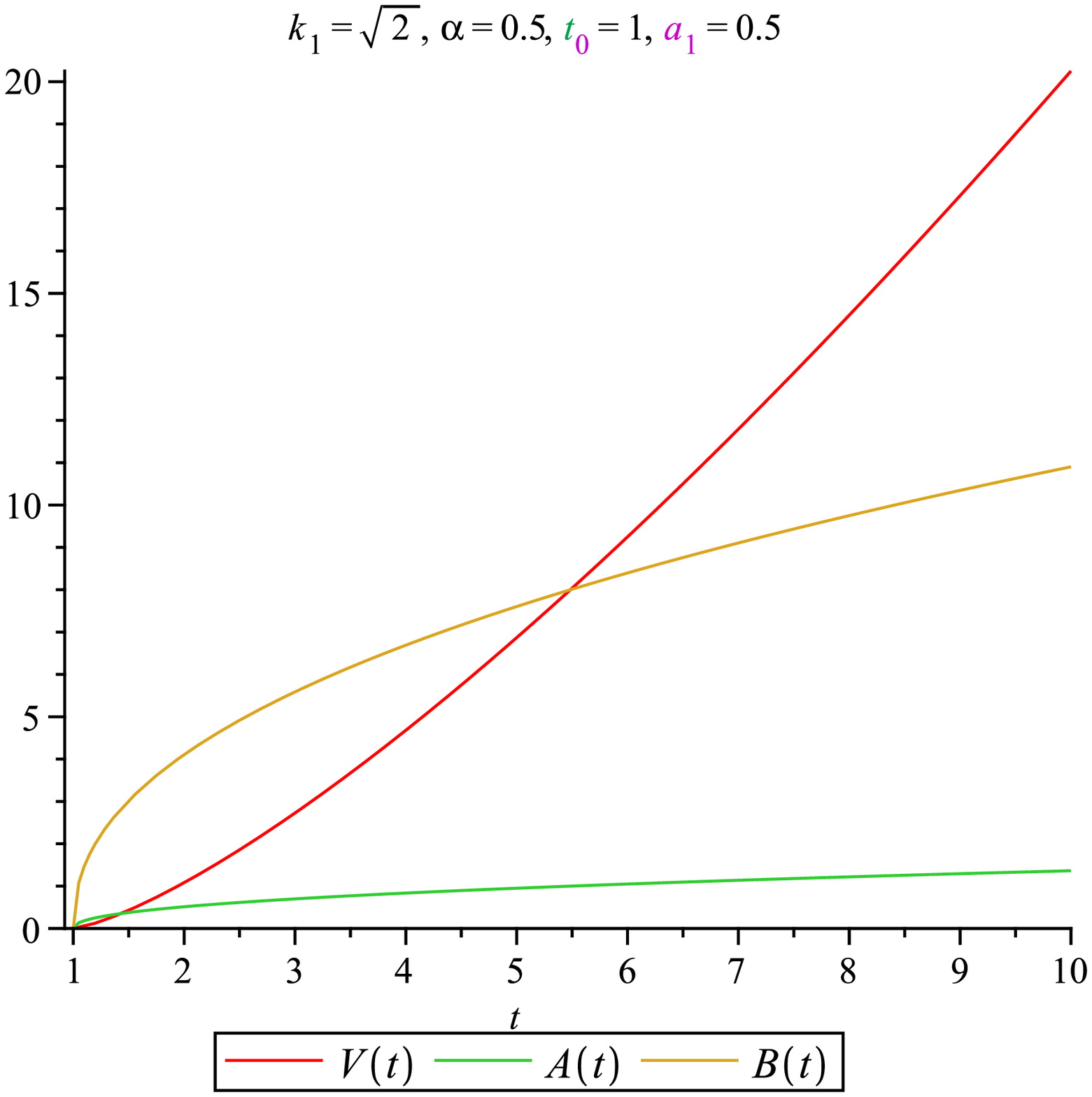}
\includegraphics[width=0.43\textwidth]{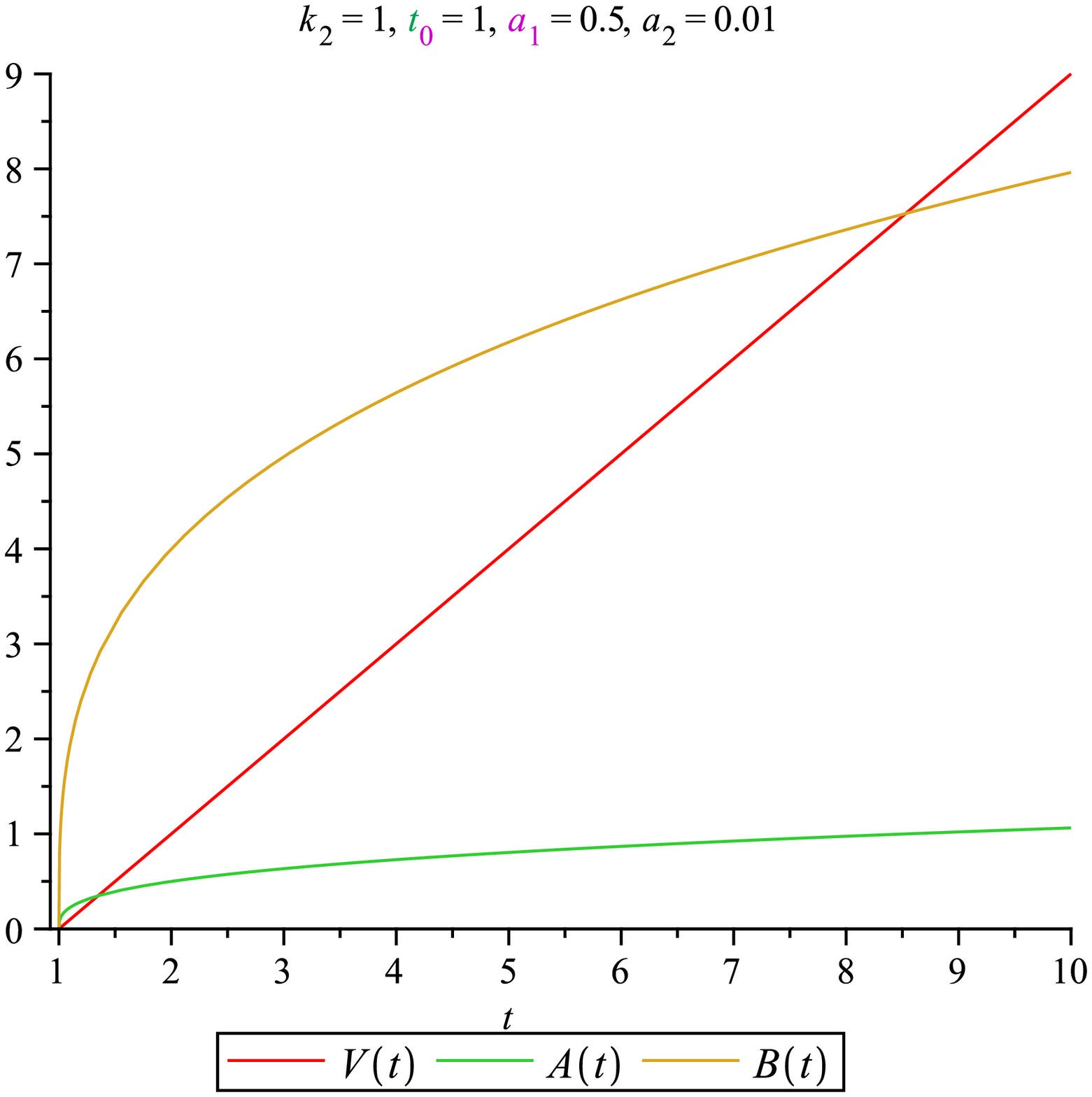}
\includegraphics[width=0.43\textwidth]{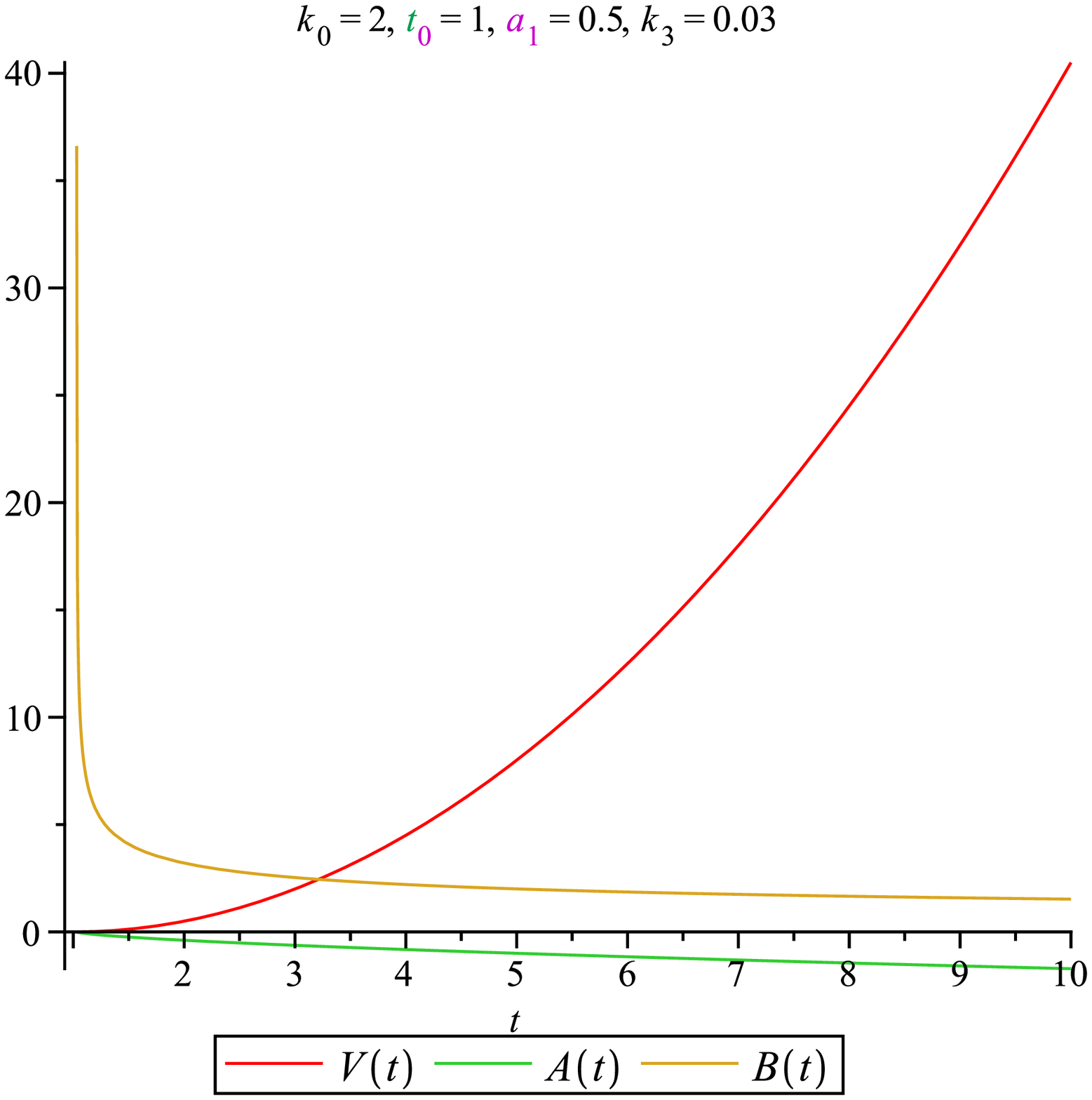}
\includegraphics[width=0.43\textwidth]{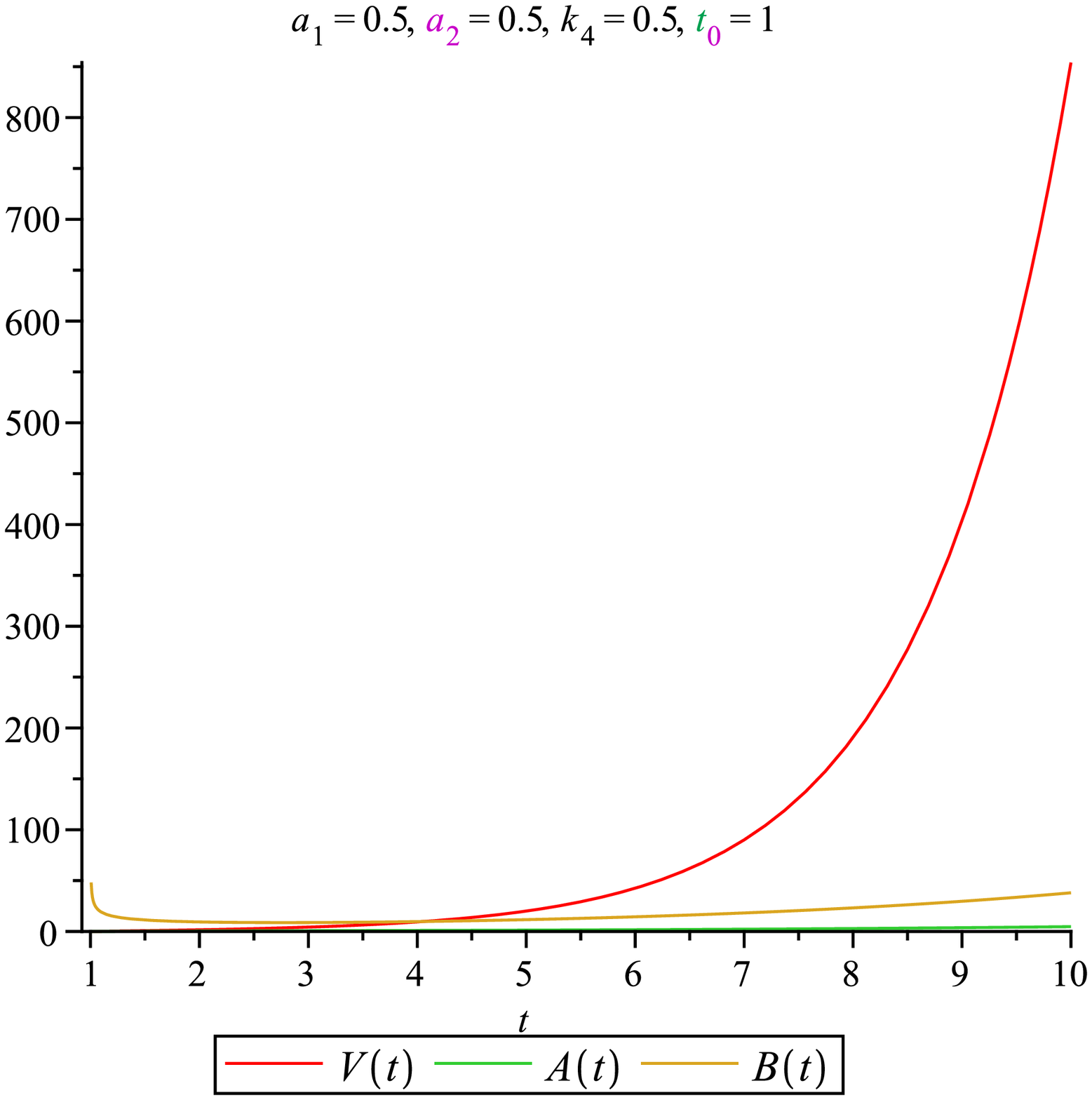}
\caption{Variation of volume $V$ and scale factors $A$ and $B$:
upper left panel for Sub-case 4.1.1 when $\rho_0 = 0$ and $a_2 =
0$, upper right panel for Sub-case 4.1.2 (i) when $\rho_0 =
0$ and $a_2 \neq 0$, $\alpha = 0$, lower left panel for
Sub-case 4.1.2 (ii) when $\rho_0 = 0$ and $a_2 \neq 0$, $\alpha =
1/\sqrt 2$, and lower right panel for Sub-case 4.1.2 (iii) when $\rho_0 = 0$ and
$a_2 \neq 0$, $\alpha = 1$,}
\end{center}
\end{figure*}

(ii) For $\alpha=\frac{1}{\sqrt{2}}$ case we can obtain the
following solution:

\begin{equation}  \label{u317-2}
\begin{array}{ll}
V(t)=\frac{k_0}{4}\,\Big(T^2-k_3^2\Big),\,\,\,\,\,\,\,\,\,\,\,\,\,\,\,
\rho(t)=\frac{1}{6\,\pi\,(1-\gamma)\,\Big(T^2-k_3^2\Big)},\\
\\
p(t)=\frac{\gamma}{6\,\pi\,(1-\gamma)\,\Big(T^2-k_3^2\Big)},\\
\\
C(t)=C_0+\frac{1}{f}\,\sqrt{\frac{\gamma}{3\,\pi\,(1-\gamma)}}\,\ln\Big[2\,\Big(T+\sqrt{T^2-k_3^2}\Big)\Big],\\
\\
A(t)=-a_1\,\,\Big(\frac{k_0}{4}\Big)^{1/3}\Big(T-k_3\Big)^{2/3},\\
\\
B(t)=\frac{1}{a_1^2}\,\Big(\frac{k_0}{4}\Big)^{1/3}\,\Big(T+k_3\Big)\,\Big(T-k_3\Big)^{-1/3},
\end{array}
\end{equation}
where $C_0$ is an arbitrary constant,
$k_3^2=\frac{36\,a_2^2}{k_0^2}$ and $T=t-t_0$.

\begin{figure*}[thbp]
\begin{center}
\vspace{0.5cm}
\includegraphics[width=0.43\textwidth]{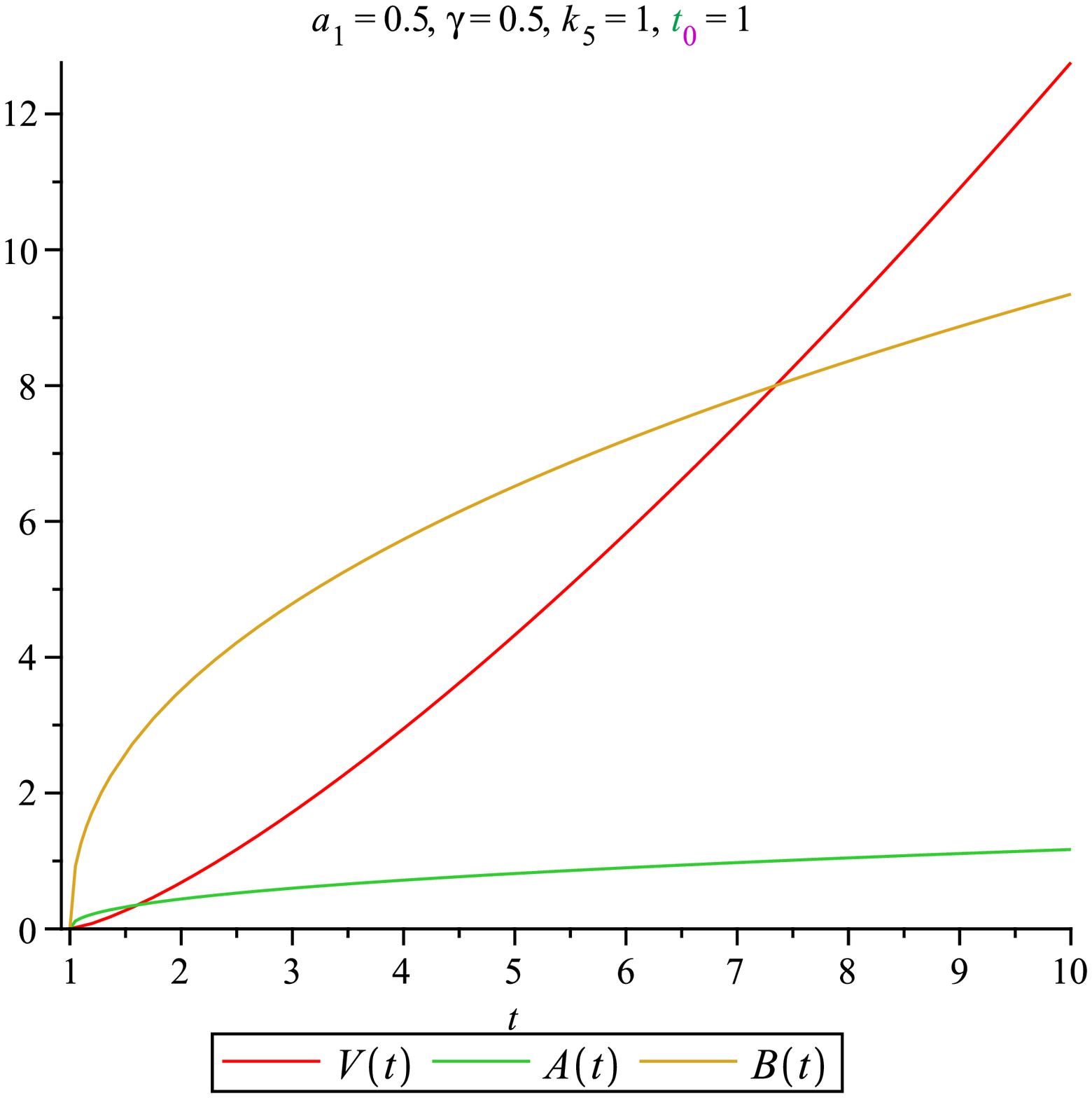}
\includegraphics[width=0.43\textwidth]{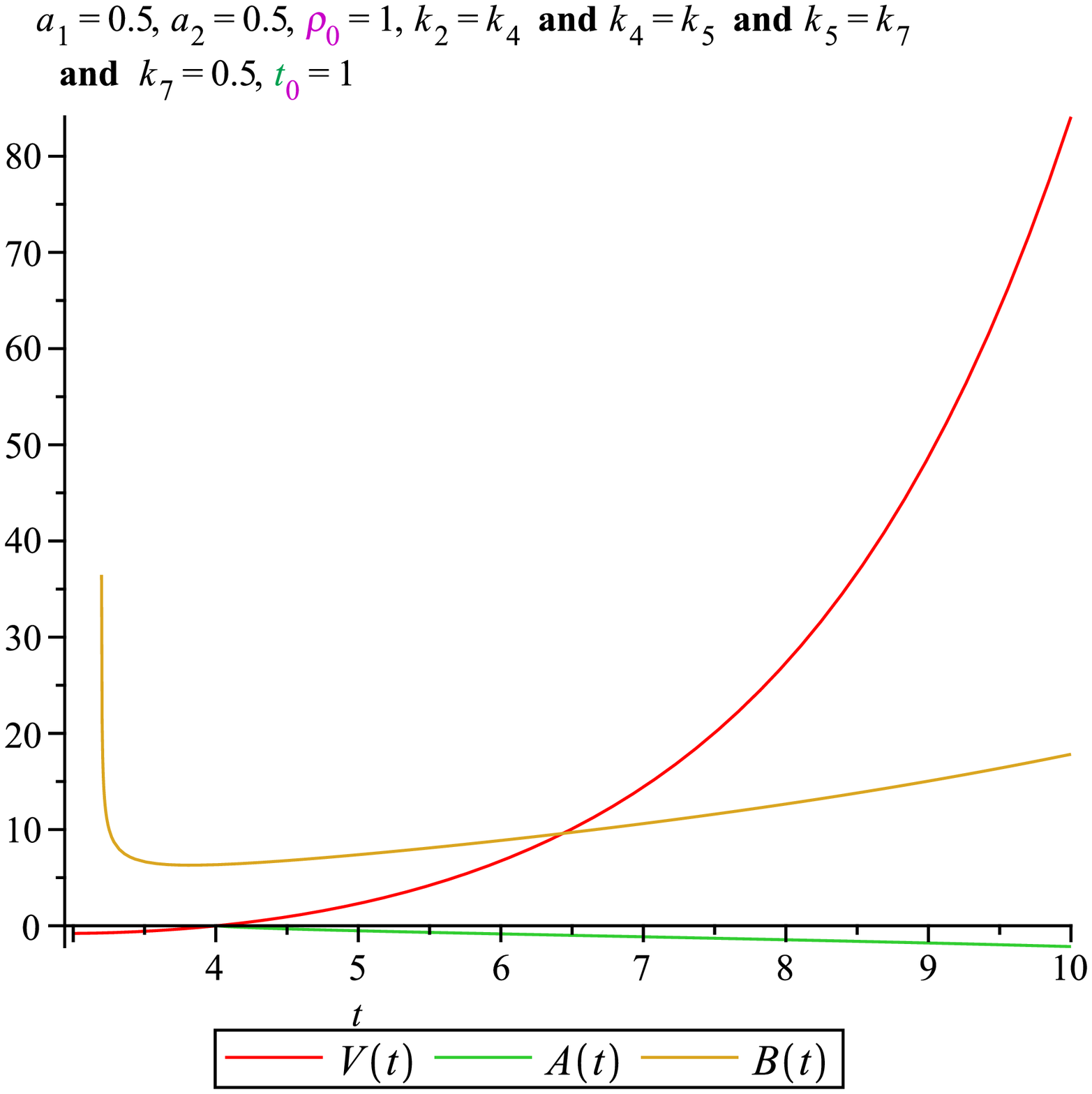}
\includegraphics[width=0.43\textwidth]{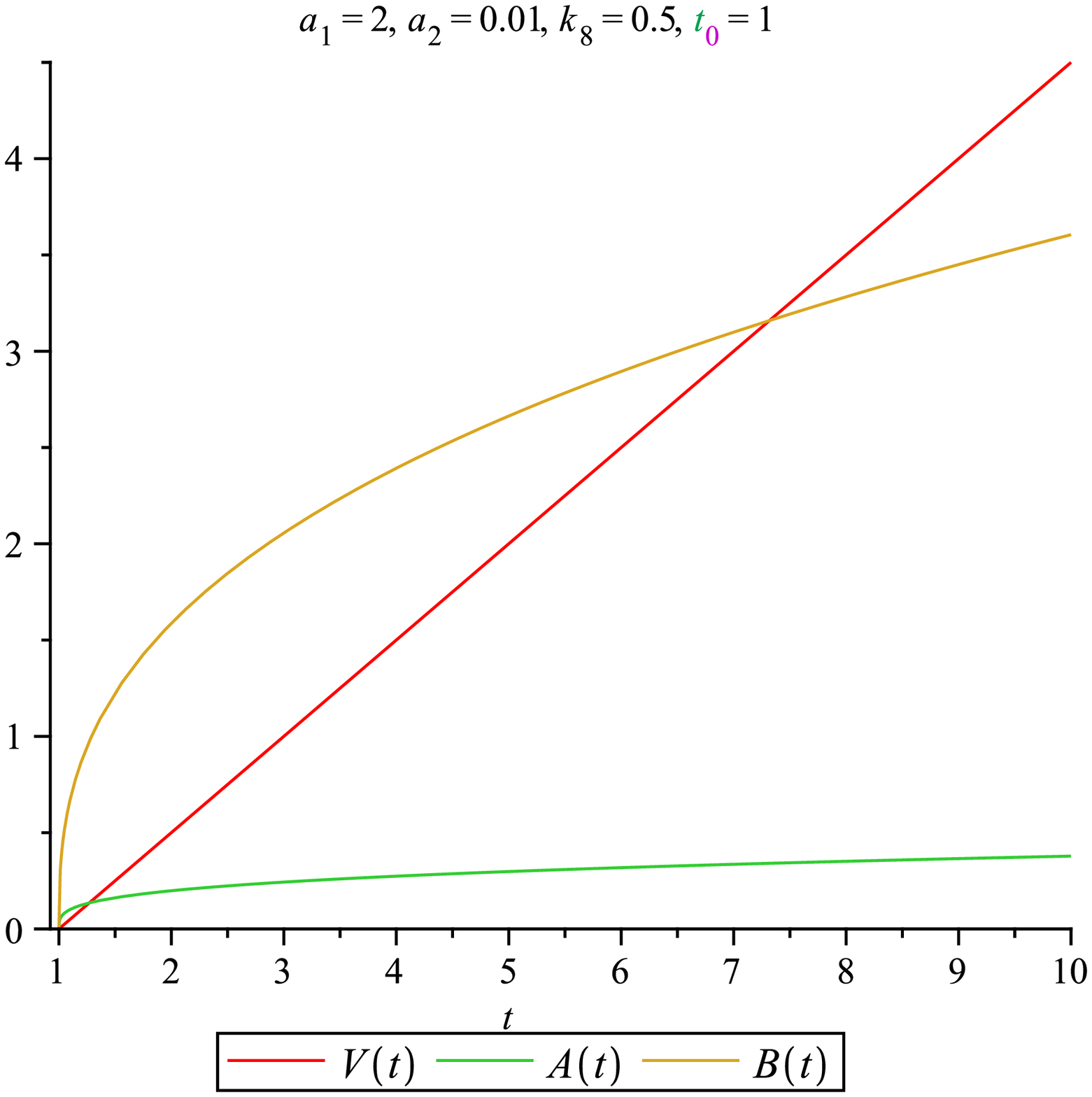}
\caption{Variation of volume $V$ and scale factors $A$ and $B$:
upper left panel for Sub-case 4.2.1 when $\rho_0 \neq 0$ and
$a_2 = 0$, upper right panel for Sub-case 4.2.2(ii) when $\rho_0 \neq 0$, $a_2 \neq 0$, $\gamma = 0$, $p(t) = 0$
and $\alpha = 1$, and lower panel for Sub-case 4.2.3 when $\rho_0
\neq 0$, $a_2 \neq 0$, $\gamma = 1$}
\end{center}
\end{figure*}

(iii) For $\alpha=1$ case we can obtain the following solution:

\begin{equation}  \label{u317-3}
\begin{array}{ll}
V(t)=k_4^{-1}\sinh\big[3\,a_2\,k_4\,T\big],\,\,\,\,\,\,\,\,\,\,\,\,\,\,\,\rho(t)=\frac{3\,a_2^2\,k_4^2}{4\,\pi\,(1-\gamma)},\\
\\
p(t)=\frac{3\,\gamma\,a_2^2\,k_4^2}{4\,\pi\,(1-\gamma)},\,\,\,\,\,\,\,\,\,\,\,\,\,\,\,
C(t)=C_0+\frac{a_2\,k_4\,T}{2\,f}\,\sqrt{\frac{3\,(1+\gamma)}{\pi\,(1-\gamma)}},\\
\\
A(t)=a_1\,k_4^{-1/3}\,\tanh^{1/3}\Big[\frac{3\,a_2\,k_4\,T}{2}\Big]\,\sinh^{1/3}\big[3\,a_2\,k_4\,T\big],\\
\\
B(t)=a_1^{-2}\,k_4^{-1/3}\,\coth^{2/3}\Big[\frac{3\,a_2\,k_4\,T}{2}\Big]\,\sinh^{1/3}\big[3\,a_2\,k_4\,T\big],
\end{array}
\end{equation}
where $C_0$ is an arbitrary constant, $k_0=2\,a_2^2\,k_4^2$ and $T=t-t_0$.\\

\begin{figure}
\begin{center}
\includegraphics[width=0.43\textwidth]{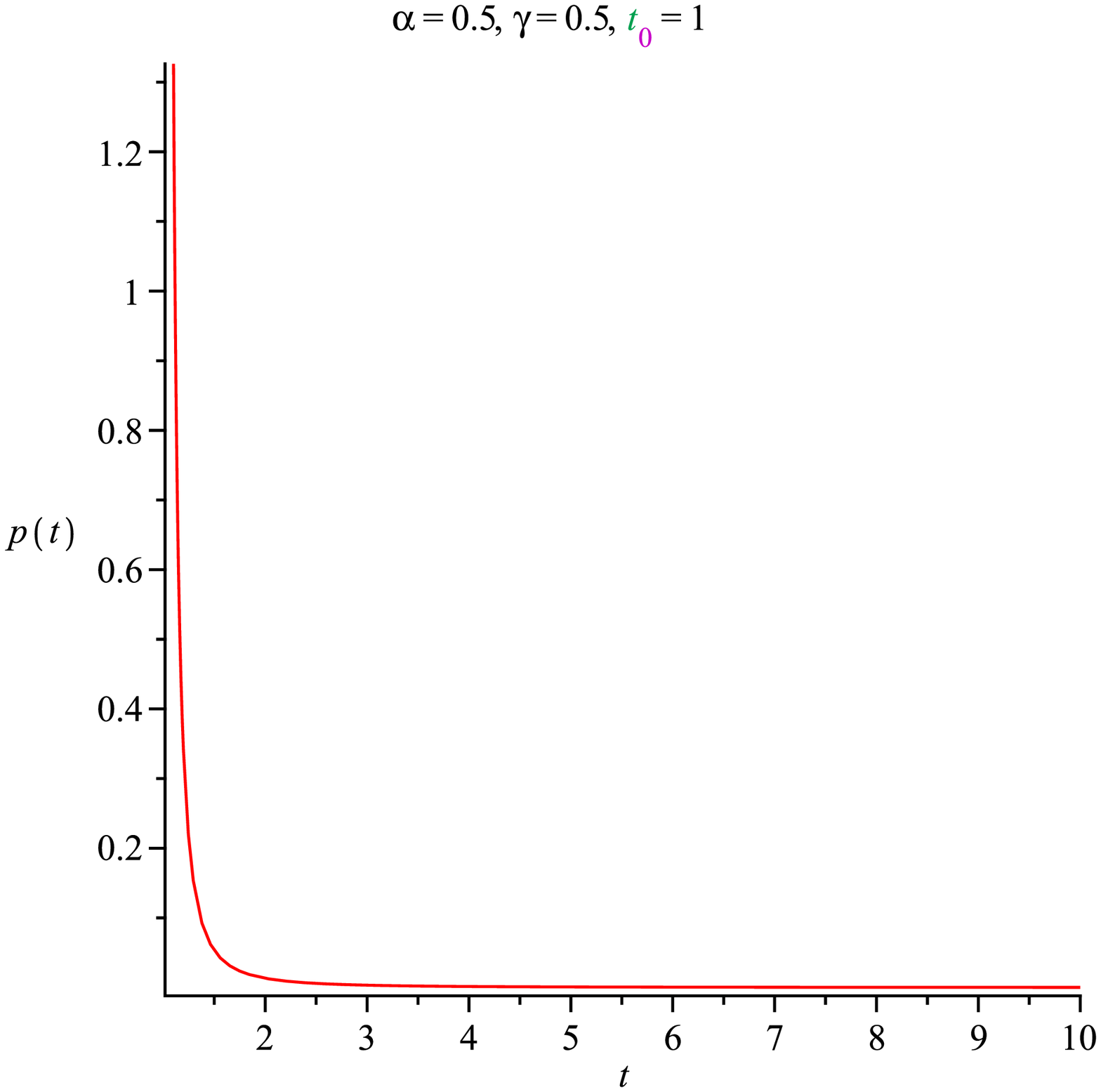}
\includegraphics[width=0.43\textwidth]{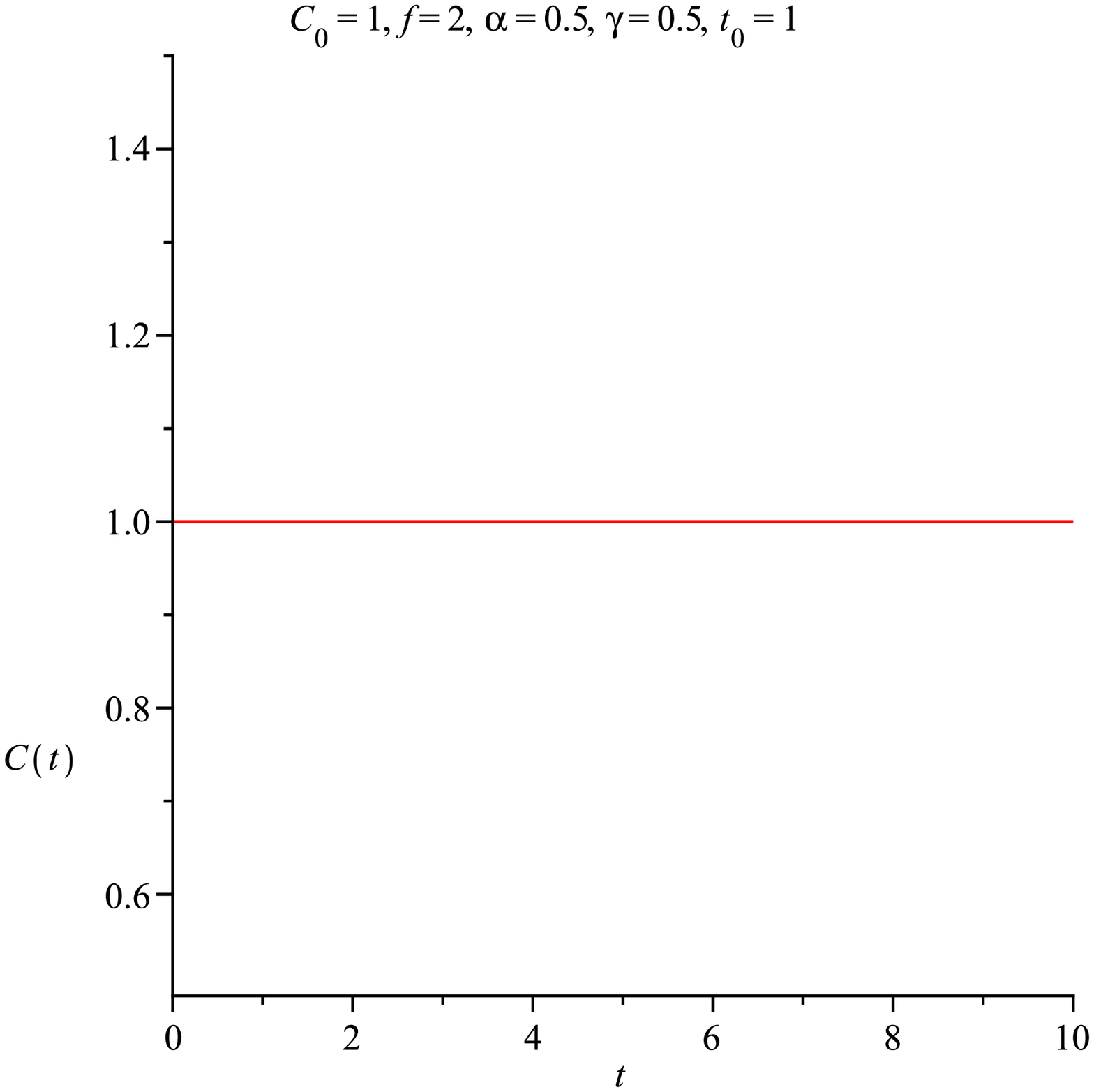}
\includegraphics[width=0.43\textwidth]{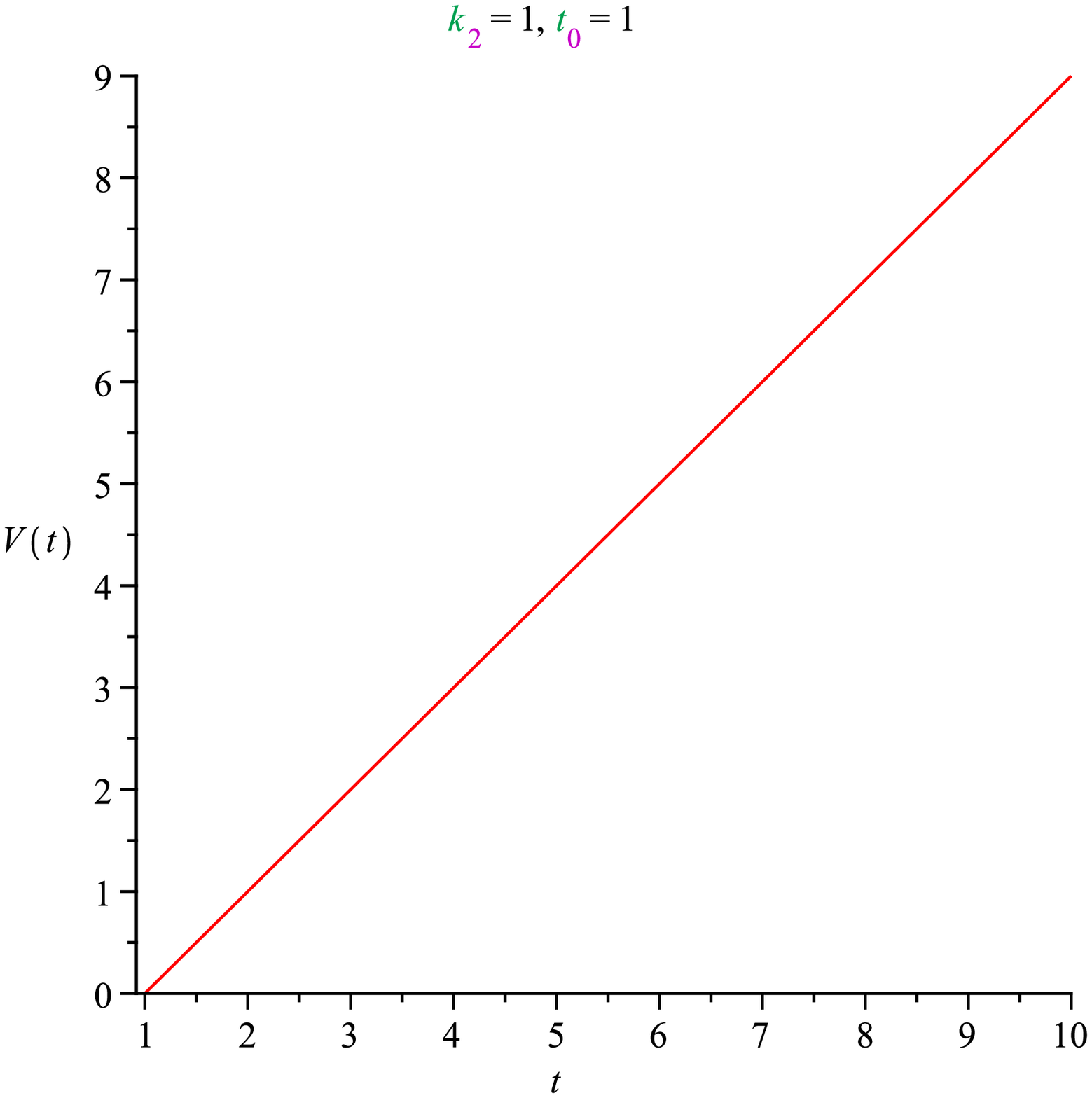}
\includegraphics[width=0.43\textwidth]{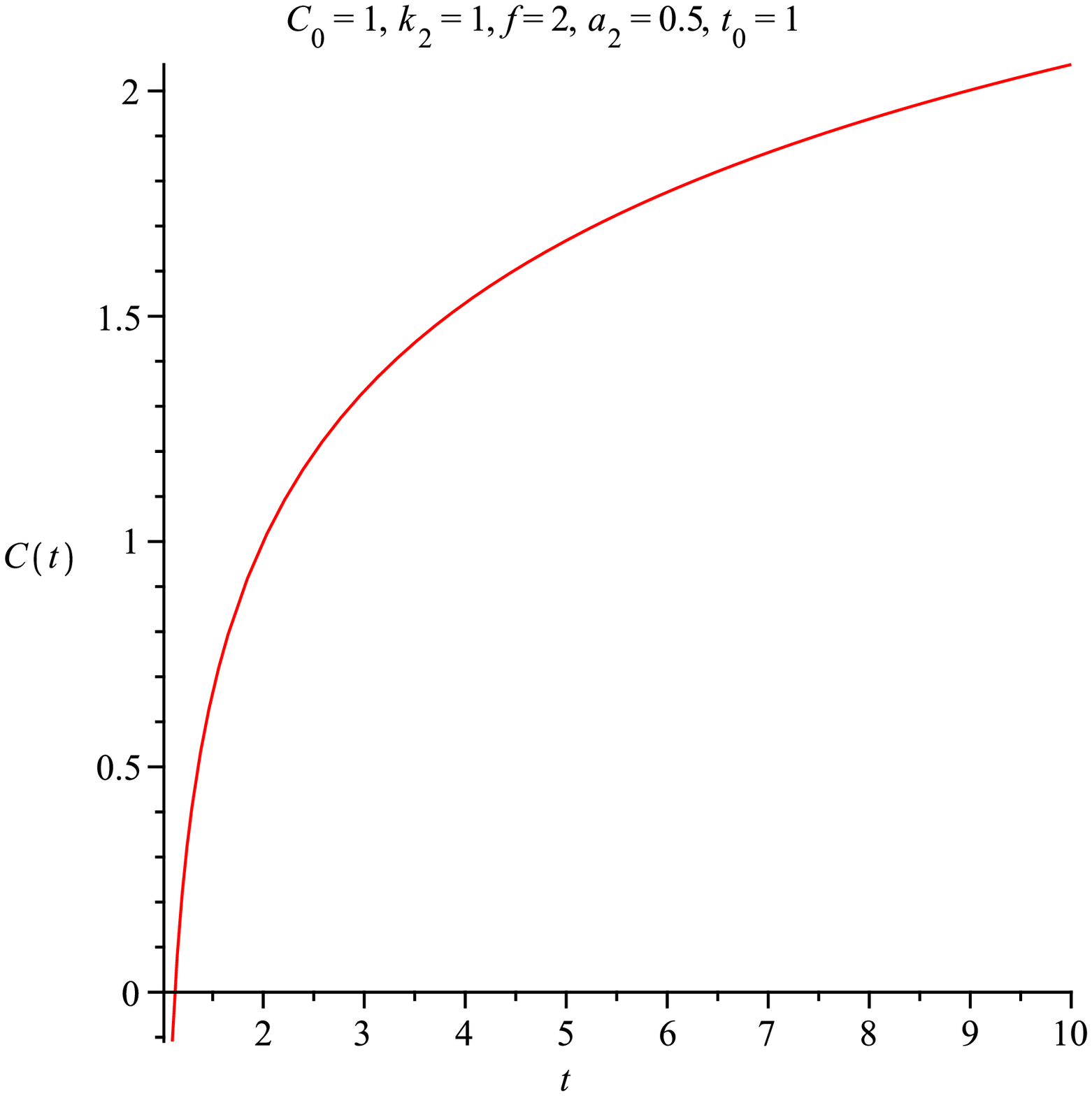}
\caption{Variation of pressure (upper left panel) and creation
field (upper right panel) for Sub-case 4.1.1 when $\rho_0 = 0$ and
$a_2 = 0$ ) whereas variation of volume (lower left panel) and
creation field (lower right panel) for Sub-case 4.1.2 (i) when
$\rho_0 = 0$ and $a_2 \neq 0$, $\alpha = 0$}
\end{center}
\end{figure}

\subsection{$\rho_0 \neq 0$}

\subsubsection{$a_2=\Omega_0=0$}

We can obtain the following solution:

\begin{equation}  \label{u318}
\begin{array}{ll}
V(t)=\Big[\frac{k_5\,(1+\gamma)\,T}{2}\Big]^{\frac{2}{1+\gamma}},\,\,\,\,\,\,\,
\rho(t)=\frac{1}{6\,\pi\,(1+\gamma)^2\,T^2},\\
\\
p(t)=\frac{\gamma}{6\,\pi\,(1+\gamma)^2\,T^2},
\,\,\,\,\,\,\,\,\,\,\,\,\,\,\,\,\,\,\,\,\,\,\,\,\,\,\,\,\,\,\,\,\,\,\,\,\,\,\,
C(t)=C_0,\\
\\
A(t)=a_1\,\Big[\frac{k_5\,(1+\gamma)\,T}{2}\Big]^{\frac{2}{3(1+\gamma)}},\,\,\,\,\,\,\,
B(t)=\frac{1}{a_1^2}\,\Big[\frac{k_5\,(1+\gamma)\,T}{2}\Big]^{\frac{2}{3(1+\gamma)}},
\end{array}
\end{equation}
where $C_0$ is an arbitrary constant, $k_5^2=3\,\rho_0$ and
$T=t-t_0$.

\begin{figure*}[thbp]
\begin{center}
\includegraphics[width=0.43\textwidth]{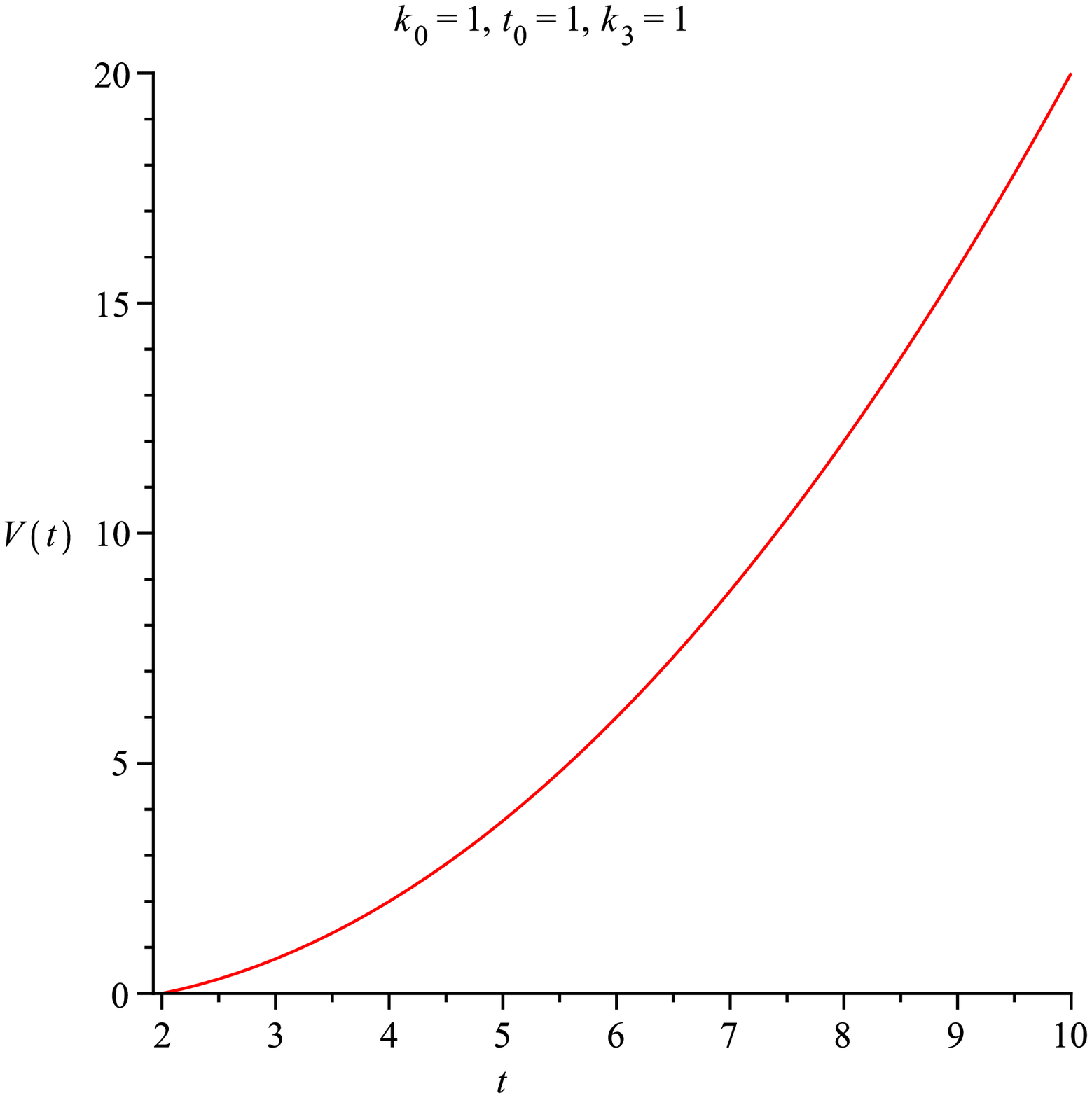}
\includegraphics[width=0.43\textwidth]{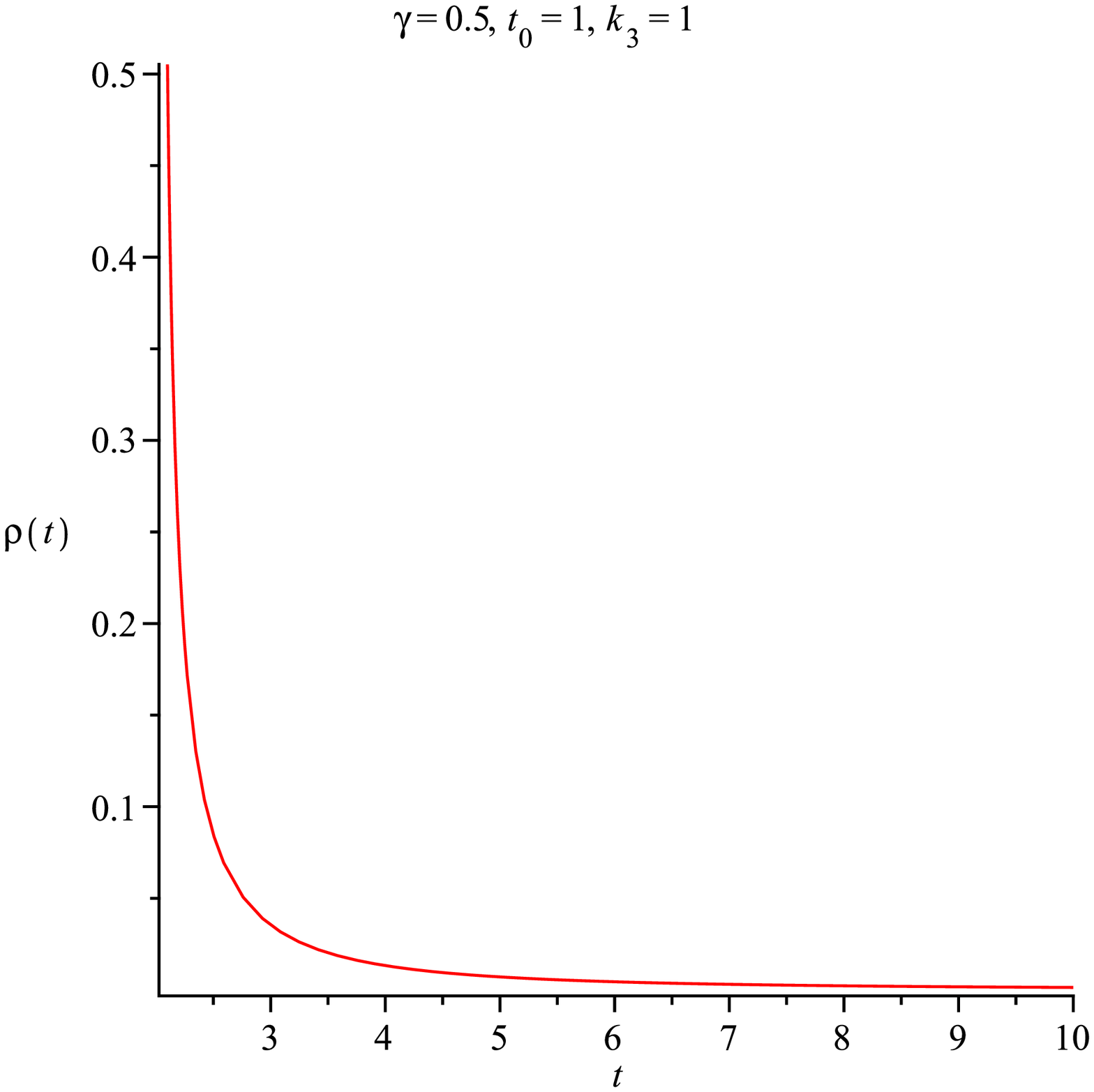}
\includegraphics[width=0.43\textwidth]{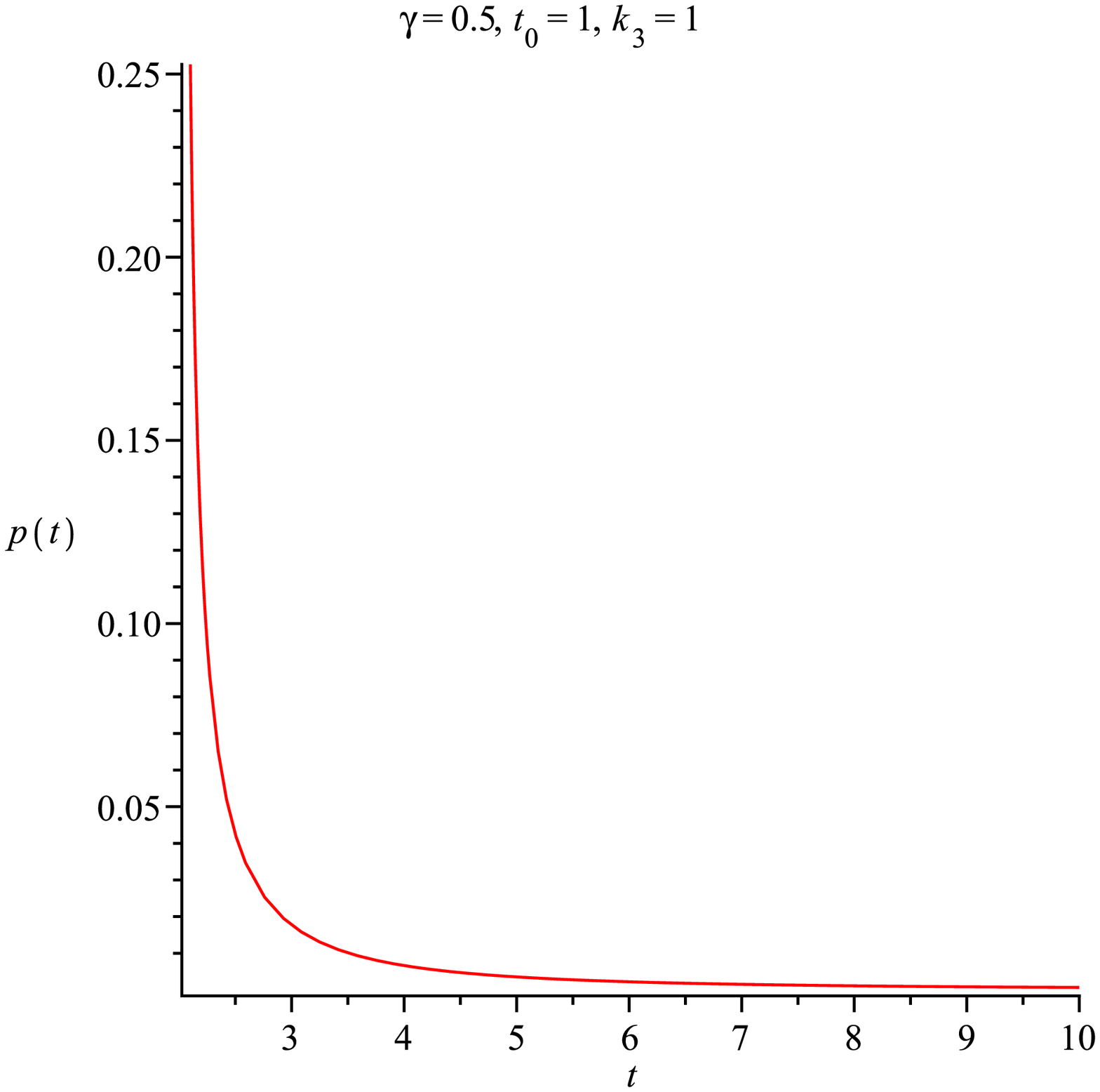}
\includegraphics[width=0.43\textwidth]{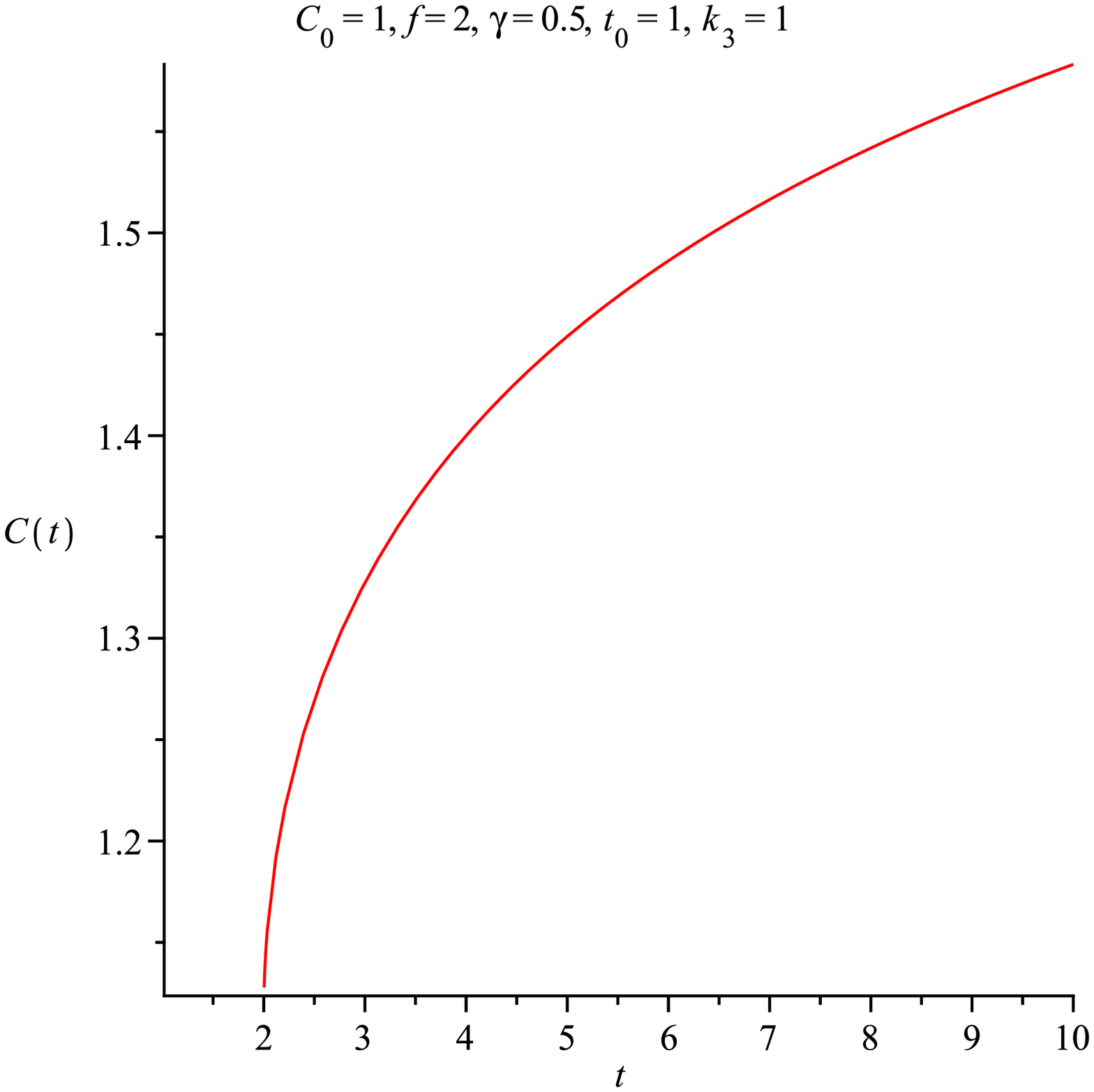}
\caption{Variation of volume, density, pressure and creation field
for Sub-case 4.1.2 (ii) when $\rho_0 = 0$, $a_2 \neq 0$, $\alpha =
1/\sqrt 2$}
\end{center}
\end{figure*}

\subsubsection{$a_2\neq 0$} When $\Omega_0\neq0$,
$\gamma=0$ and $p(t)=0$.

(i) For $\alpha=0$ case we can obtain the following solution:

\begin{equation}  \label{u319-1}
\begin{array}{ll}
V(t)=\frac{3\,\rho_0}{4}\,\Big(T^2-k_6^2\Big),\,\,\,\,\,\,\,\,\,\,\,\,\,\,\,
\rho(t)=\frac{1}{6\,\pi\,\Big(T^2-k_6^2\Big)},\\
\\
C(t)=C_0+\frac{1}{f\,\rho_0\,k_6}\sqrt{\frac{4\,a_2^2-\rho_0^2\,k_6^2}{3\,\pi}}\,\ln\Big[\frac{k_6+T}{k_6-T}\Big],\\
\\
A(t)=a_1\,\Big[\frac{3\,\rho_0\,(T^2-k_6^2)}{4}\Big]^{1/3}\,\Big[\frac{k_6-T}{k_6+T}\Big]^{\frac{2\,a_2}{3\,\rho_4\,k_6}},\\
\\
B(t)=\frac{1}{a_1^2}\,\Big[\frac{3\,\rho_0\,(T^2-k_6^2)}{4}\Big]^{1/3}\,\Big[\frac{k_6+T}{k_6-T}\Big]^{\frac{4\,a_2}{3\,\rho_0\,k_6}},
\end{array}
\end{equation}
where $C_0$ is an arbitrary constant,
$4\,\Omega_0-12\,a_2^2=-3\,\rho_0\,k_6^2$ and $T=t-t_0$.

\begin{figure*}[thbp]
\begin{center}
\includegraphics[width=0.43\textwidth]{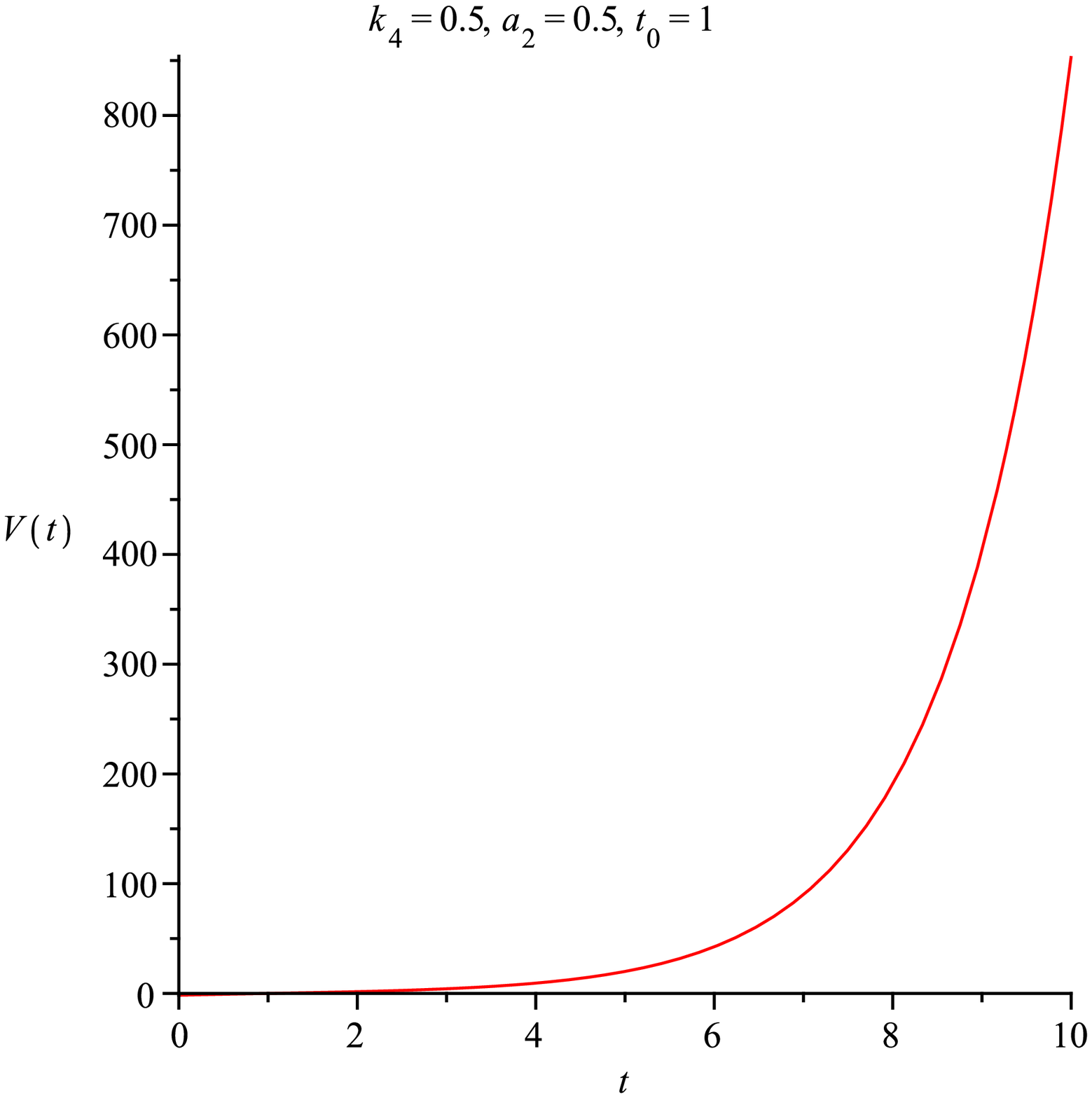}
\includegraphics[width=0.43\textwidth]{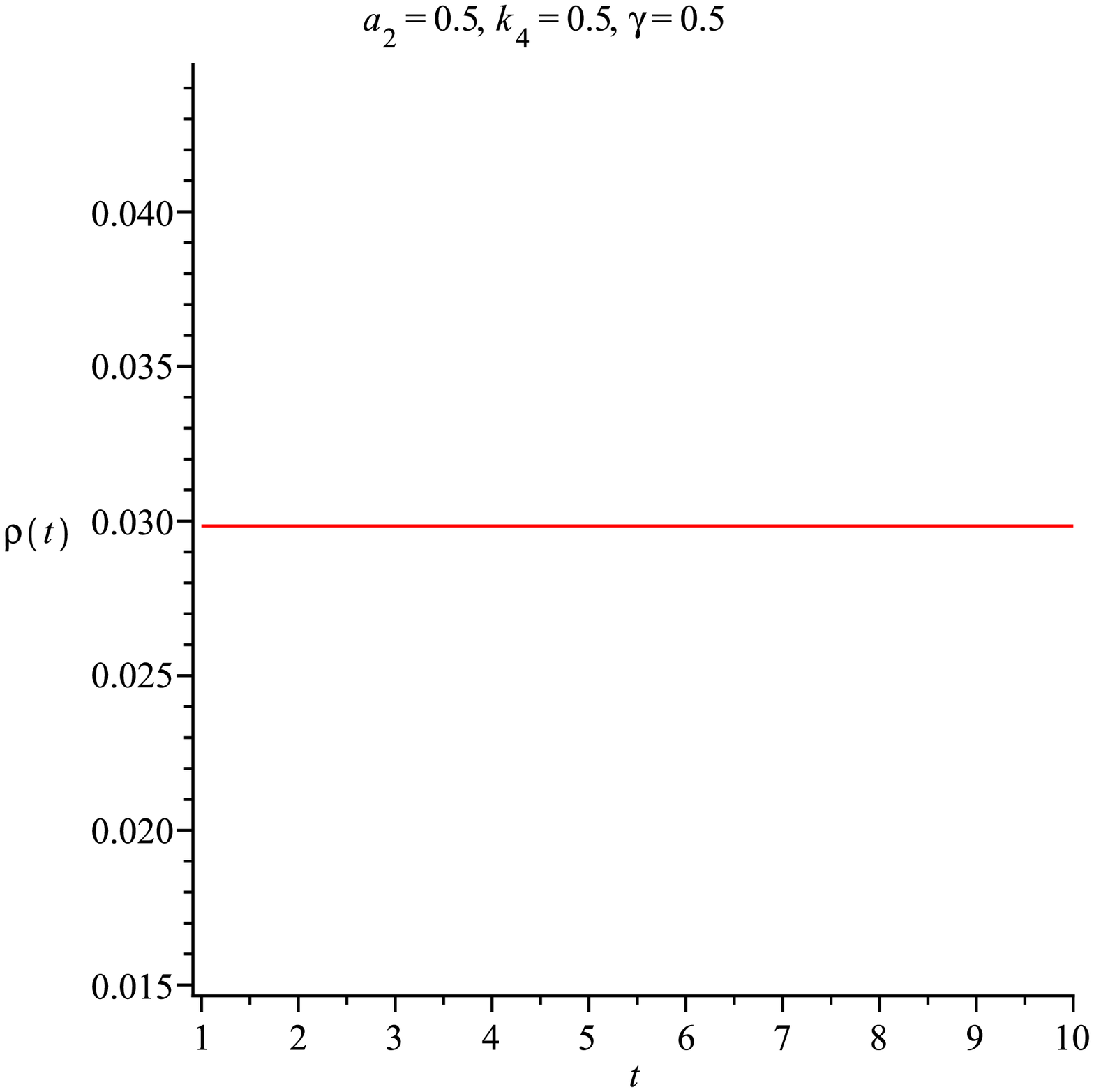}
\includegraphics[width=0.43\textwidth]{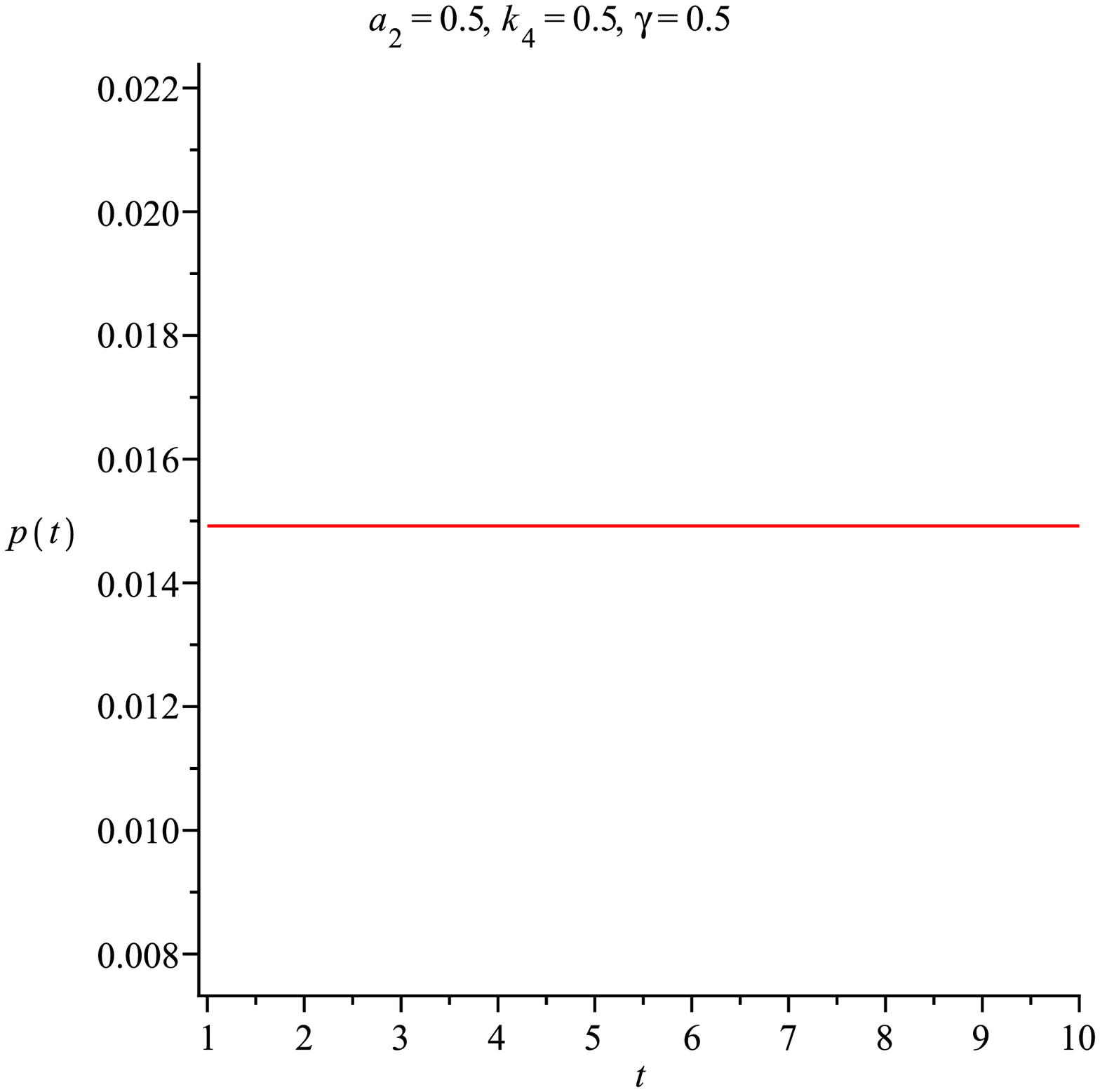}
\includegraphics[width=0.43\textwidth]{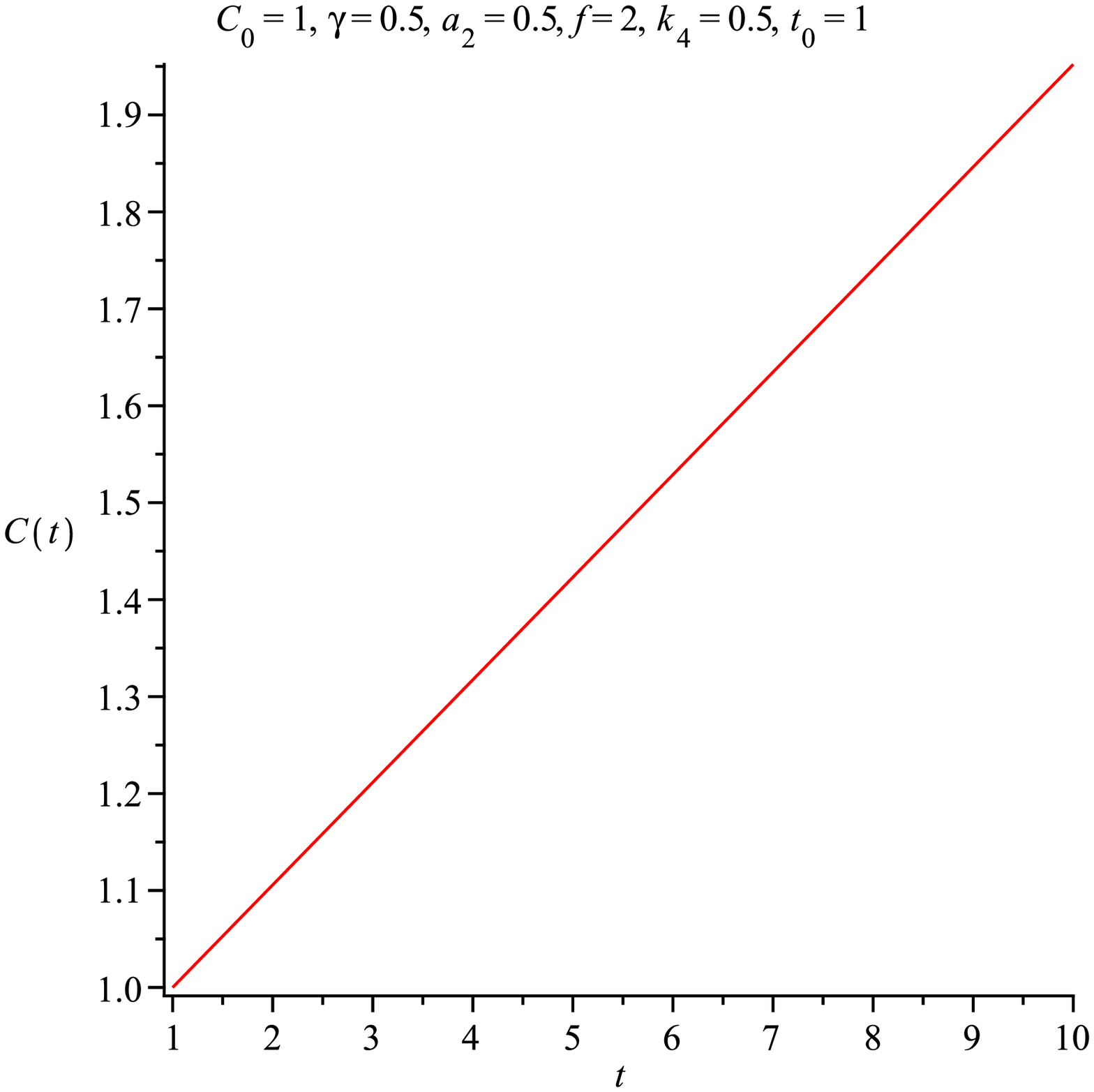}
\caption{Variation of volume, density, pressure and creation field
for Sub-case 4.1.2 (iii) when $\rho_0 = 0$, $a_2 \neq 0$, $\alpha
= 1$}
\end{center}
\end{figure*}

(ii) For $\alpha=1$ case we can obtain the following solution:

\begin{equation}  \label{u319-2}
\begin{array}{ll}
V(t)=\frac{1}{4\,k_7^2}\,\mathrm{e}^{-k_7\,T}\Big[\Big(\mathrm{e}^{k_7\,T}-3\,\rho_0\Big)^2-36\,a_2^2\,k_7^2\Big],\\
\\
\rho(t)=\frac{k_7^2}{12\,\pi}\,\Big[1-\frac{6\,\rho_0\,\mathrm{e}^{k_7\,T}}{\mathrm{e}^{2\,k_7\,T}+9\,\rho_0^2-36\,a_2^2\,k_7^2}\Big]^{-1},\\
\\
C(t)=C_0+\frac{k_7\,T}{2\,f\,\sqrt{3\,\pi}},\\
\\
A(t)=-a_1\,\Big(\frac{1}{2\,k_7}\Big)^{2/3}\,\mathrm{e}^{-\frac{k_7\,T}{3}}\,\Big[\mathrm{e}^{k_7\,T}-3\,\rho_0-6\,a_2\,k_7\Big]^{2/3},\\
\\
B(t)=\frac{1}{a_1^2}\,\Big(\frac{1}{2\,k_7}\Big)^{2/3}\,\mathrm{e}^{-\frac{k_7\,T}{3}}\,\Big[\mathrm{e}^{k_7\,T}-3\,\rho_0+6\,a_2\,k_7\Big]
\,\Big[\mathrm{e}^{k_5\,T}-3\,k_4-6\,k_2\,K_5\Big]^{-1/3},
\end{array}
\end{equation}
where $C_0$ is an arbitrary constant, $3\,\Omega_0=k_7^2$ and
$T=t-t_0$.

\begin{figure*}[thbp]
\begin{center}
\includegraphics[width=0.43\textwidth]{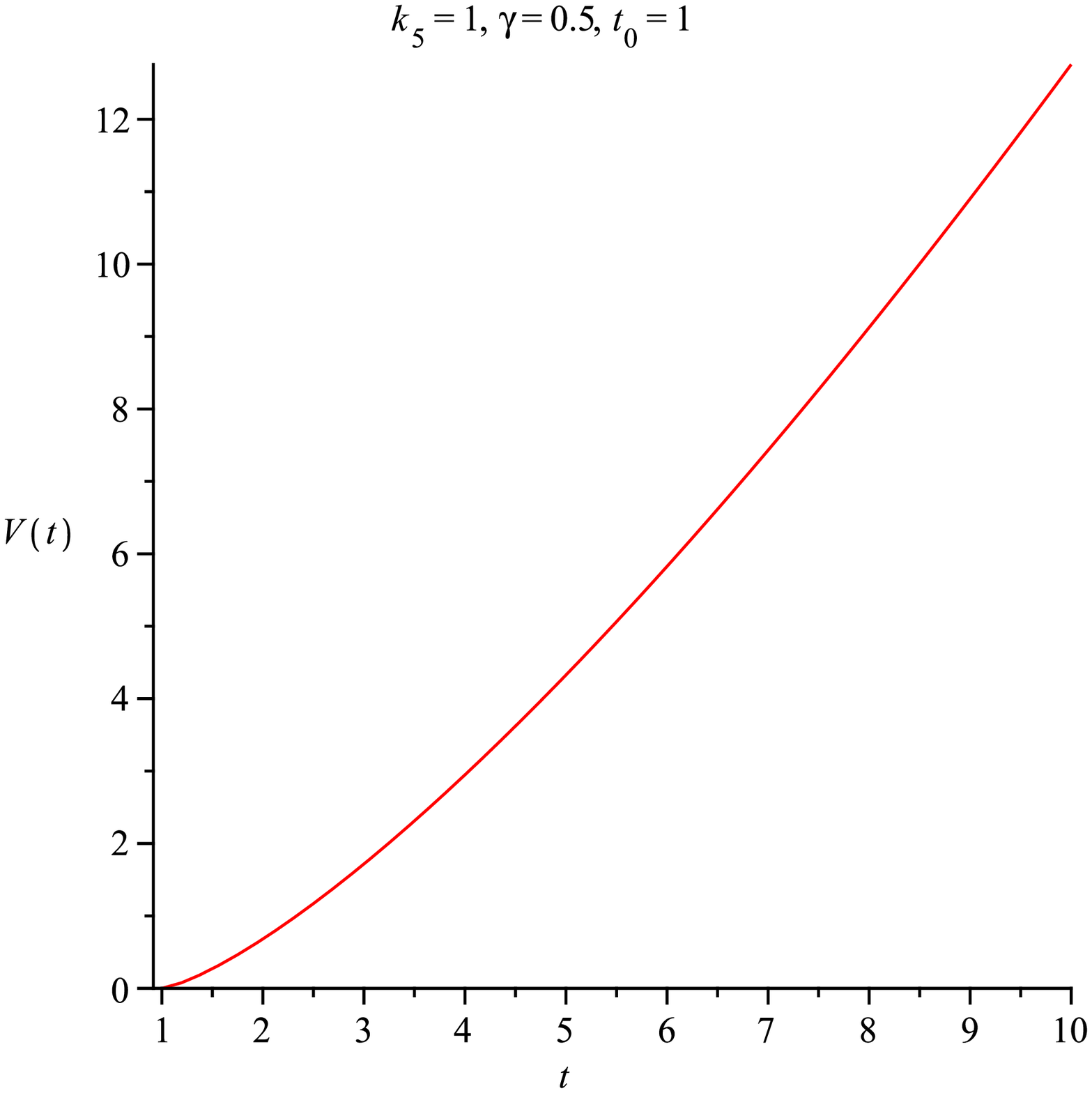}
\includegraphics[width=0.43\textwidth]{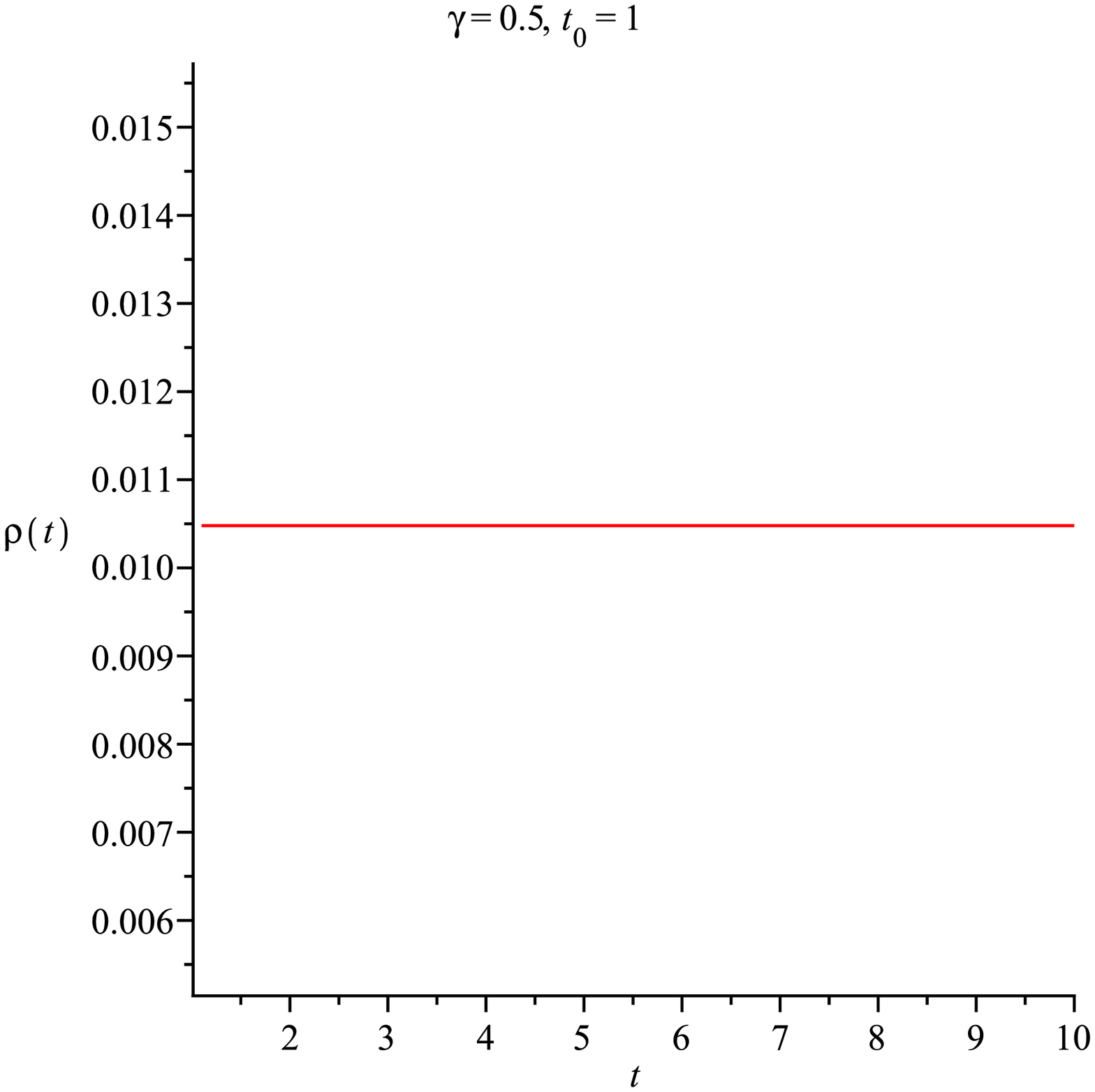}
\includegraphics[width=0.43\textwidth]{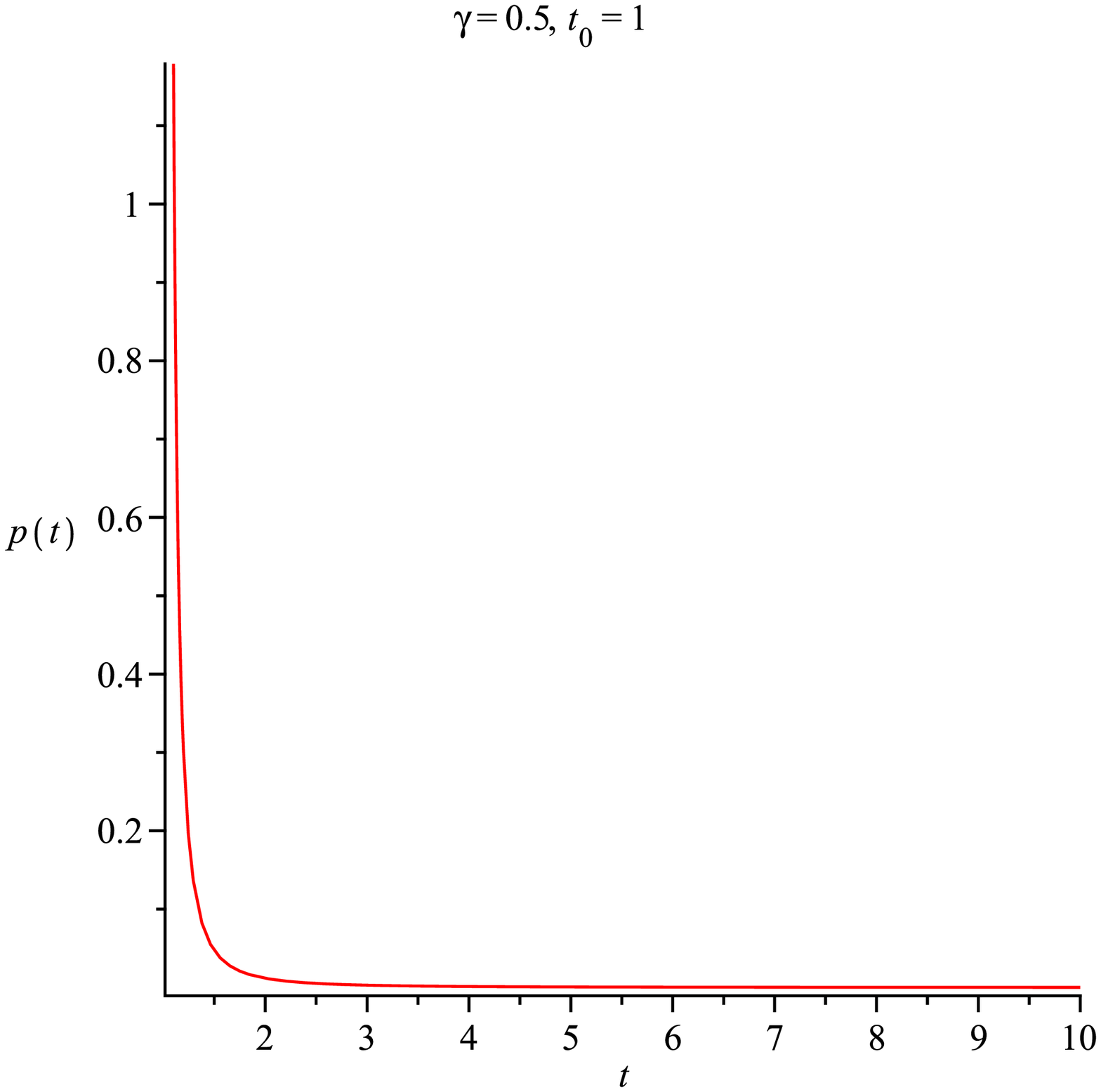}
\includegraphics[width=0.43\textwidth]{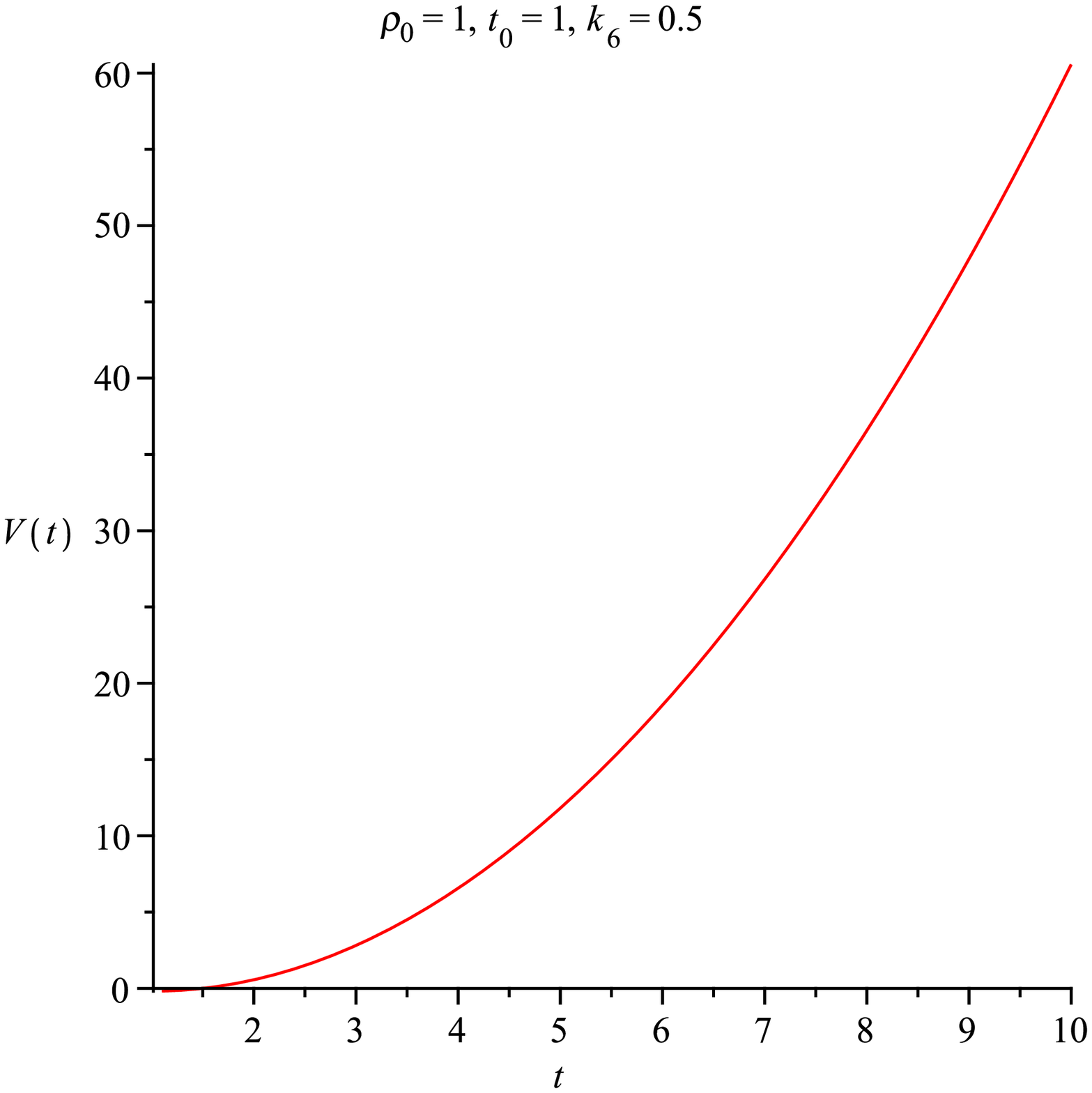}
\caption{Variation of volume (upper left panel), density (upper
right panel), pressure (lower left panel) for Sub-case 4.2.1 when
$\rho_0 \neq 0$, $a_2 = 0$, $\Omega_0 =O$, and variation of volume
(lower right panel) for Sub-case 4.2.2. (i) when $\rho_0 \neq 0$,
$a_2 \neq 0$, $\Omega_0 \neq 0$, $\gamma=0$, $p(t) = 0$ and
$\alpha =0$}
\end{center}
\end{figure*}

\begin{figure*}[thbp]
\begin{center}
\includegraphics[width=0.43\textwidth]{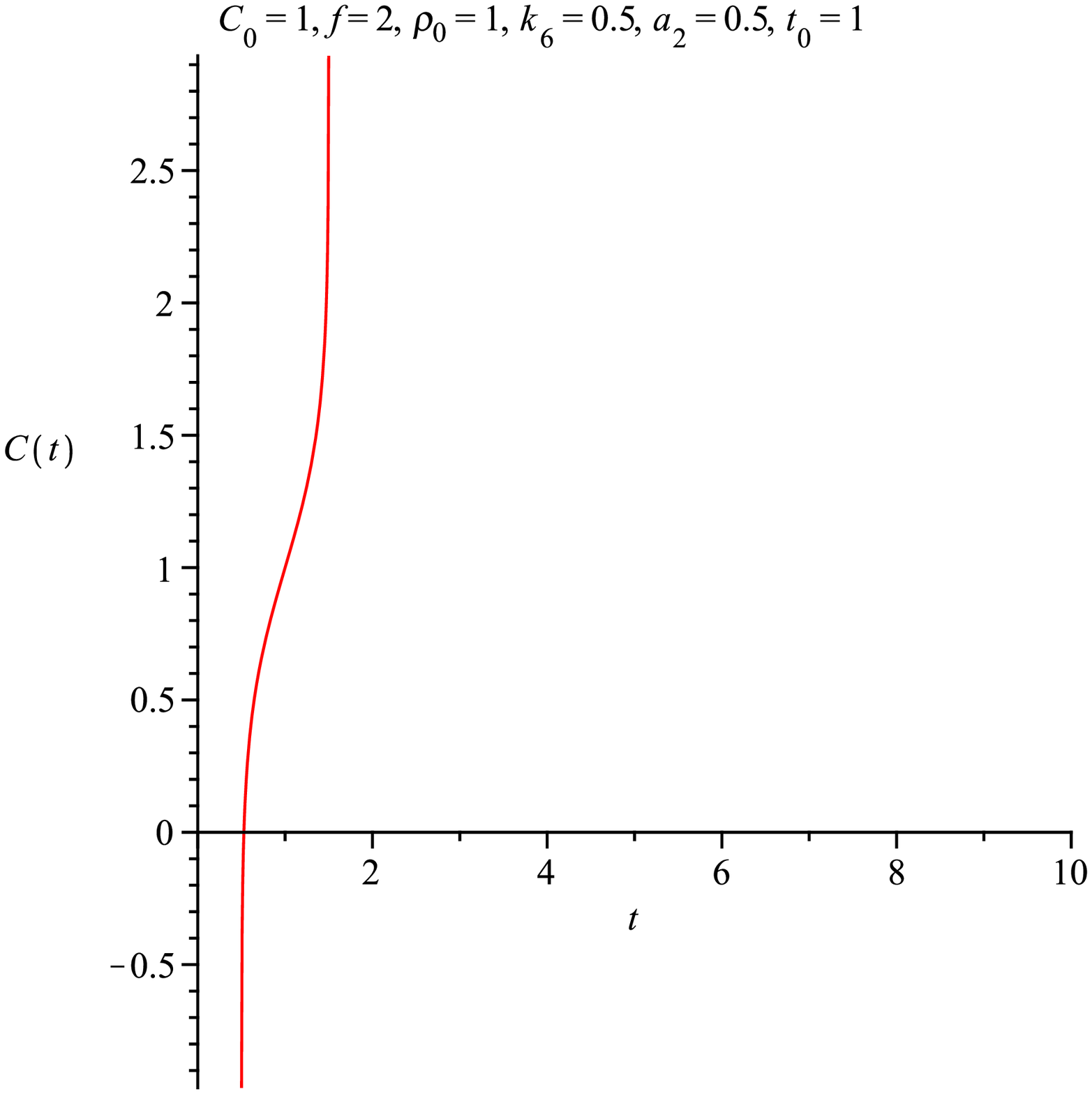}
\includegraphics[width=0.43\textwidth]{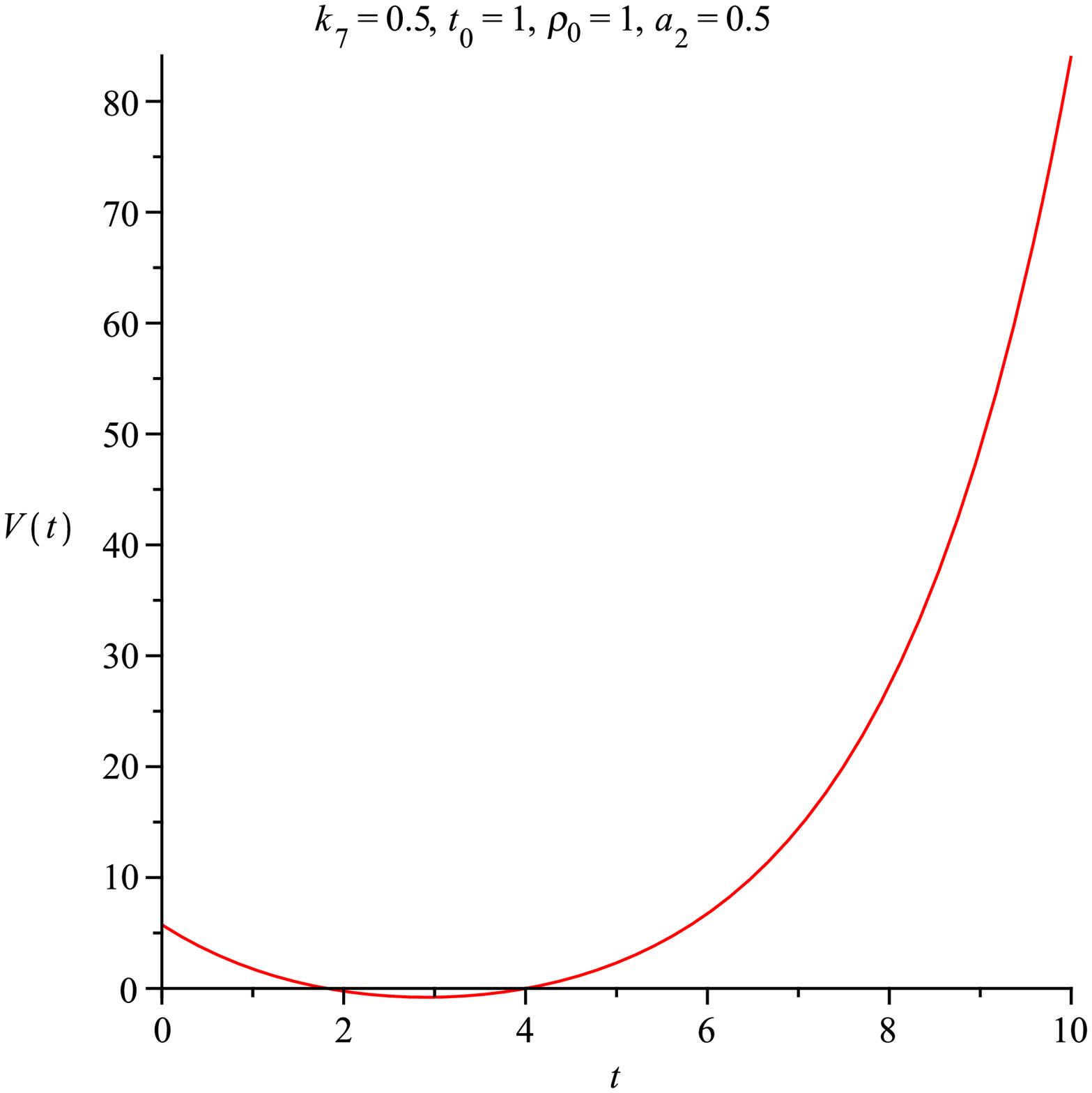}
\caption{Variation of creation field (left panel) for Sub-case
4.2.2 (i) when $\rho_0 \neq 0$, $a_2 \neq 0$, $\Omega_0 \neq 0$,
$\gamma=0$, $p(t) = 0$ and $\alpha =0$ whereas variation of volume
(right panel) for Sub-case 4.2.2 (ii) when $\rho_0 \neq 0$, $a_2
\neq 0$, $\Omega_0 \neq 0$, $\gamma=0$, $p(t) = 0$ and $\alpha
=1$}
\end{center}
\end{figure*}

\subsubsection{$a_2\neq 0$} When $\Omega_0>0$,
$3\,\rho_0+9\,a_2^2=k_8^2$ and $\gamma=1$. In this case we can
obtain the following solution:

\begin{equation}  \label{u320}
\begin{array}{ll}
V(t)=k_8\,T,\,\,\,\,\,\,\,\,\,\,
\rho(t)=p(t)=\frac{1}{24\,\pi\,k_8^2\,T^2}\,\Big[k_8^2-9\,a_2^2+3\,\rho_0\,\Big(k_8\,T\Big)^{2\,\alpha^2}\Big],\\
\\
C(t)=C_0+\frac{\sqrt{\rho_0}}{2\,f\,k_8\,\alpha^2\,\sqrt{\pi}}\,\Big(k_8\,T\Big)^{2\,\alpha^2},\\
\\
A(t)=\frac{k_8^{1/3}}{a_1^2}\,T^{\frac{1}{3}-\frac{2\,a_2}{k_8}},\,\,\,\,\,\,\,\,\,\,
B(t)=a_1\,k_8^{1/3}\,T^{\frac{1}{3}+\frac{2\,a_2}{k_8}},,
\end{array}
\end{equation}
where $C_0$ is an arbitrary constant and $T=t-t_0$.

\begin{figure*}[thbp]
\begin{center}
\includegraphics[width=0.43\textwidth]{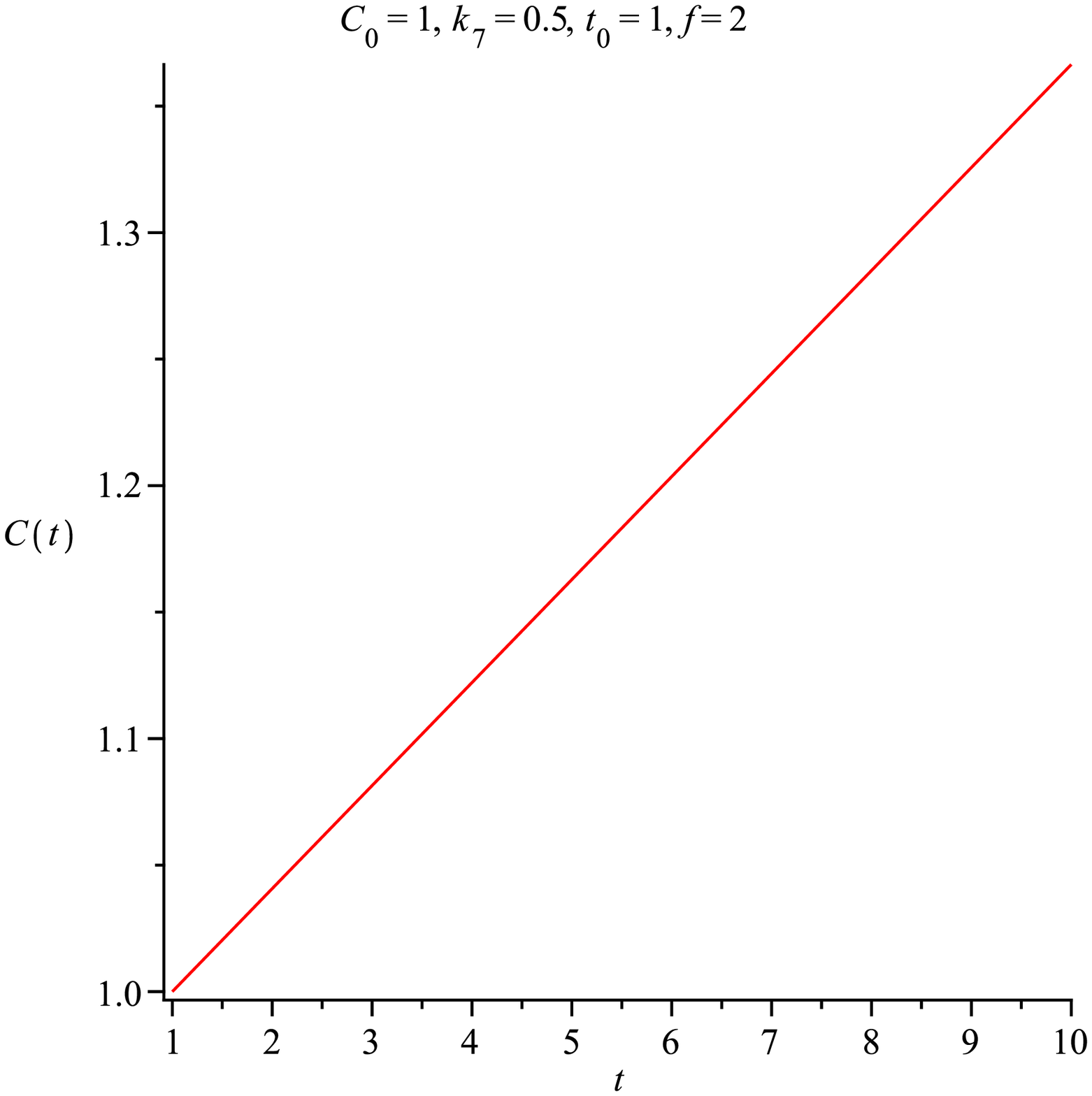}
\includegraphics[width=0.43\textwidth]{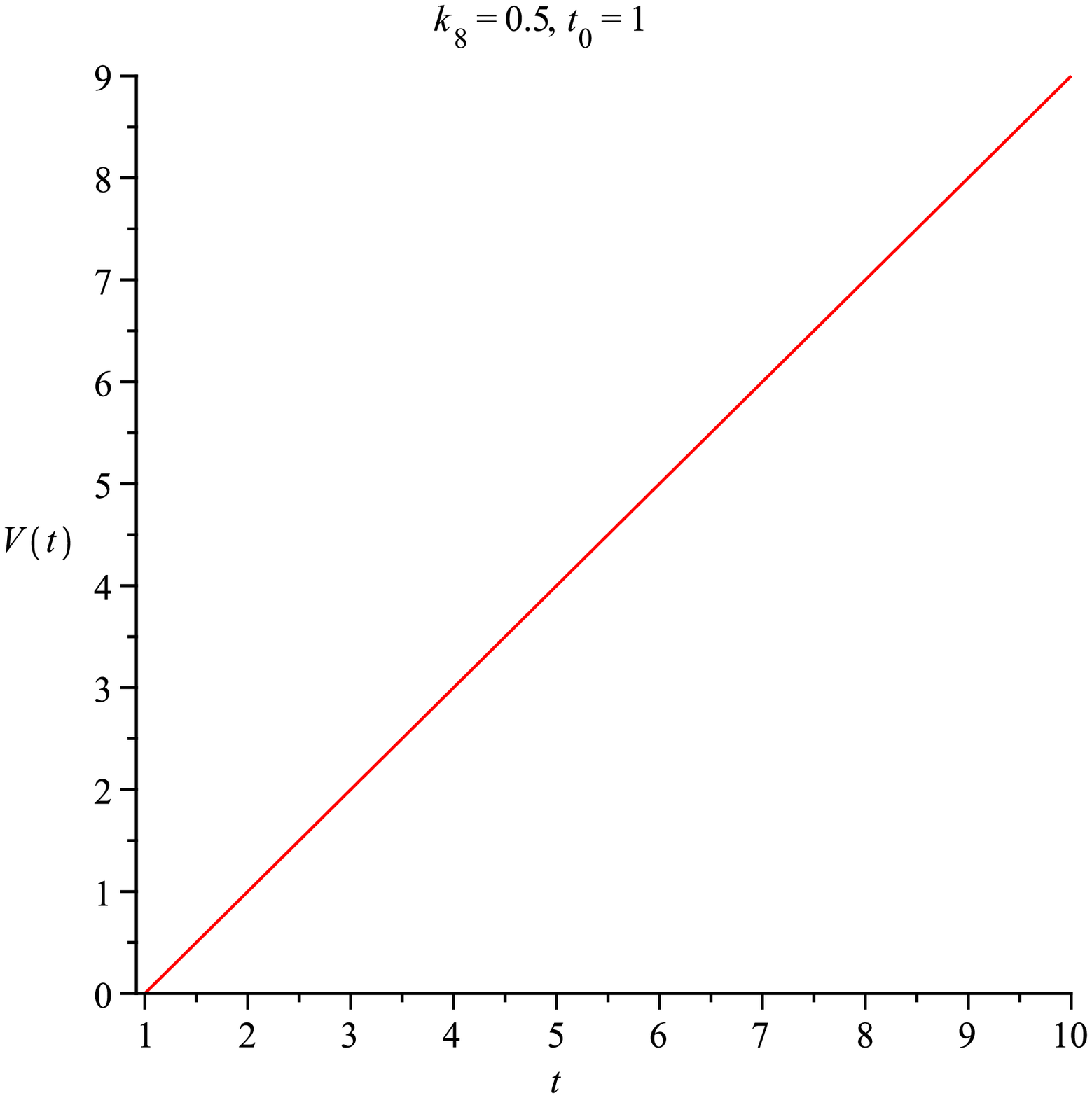}
\caption{Variation of creation field (left panel) for Sub-case
4.2.2 (ii) when $\rho_0 \neq 0$, $a_2 \neq 0$, $\Omega_0 \neq 0$,
$\gamma=0$, $p(t) = 0$ and $\alpha =1$ whereas variation of volume
(right panel) for Sub-case 4.2.3 when $\rho_0 \neq 0$, $a_2 \neq
0$, $\Omega_0> 0$, $3\rho_0+9a_2^2=k_8^2$ and $\gamma=1$}
\end{center}
\end{figure*}

\begin{figure}
\begin{center}
\includegraphics[width=0.43\textwidth]{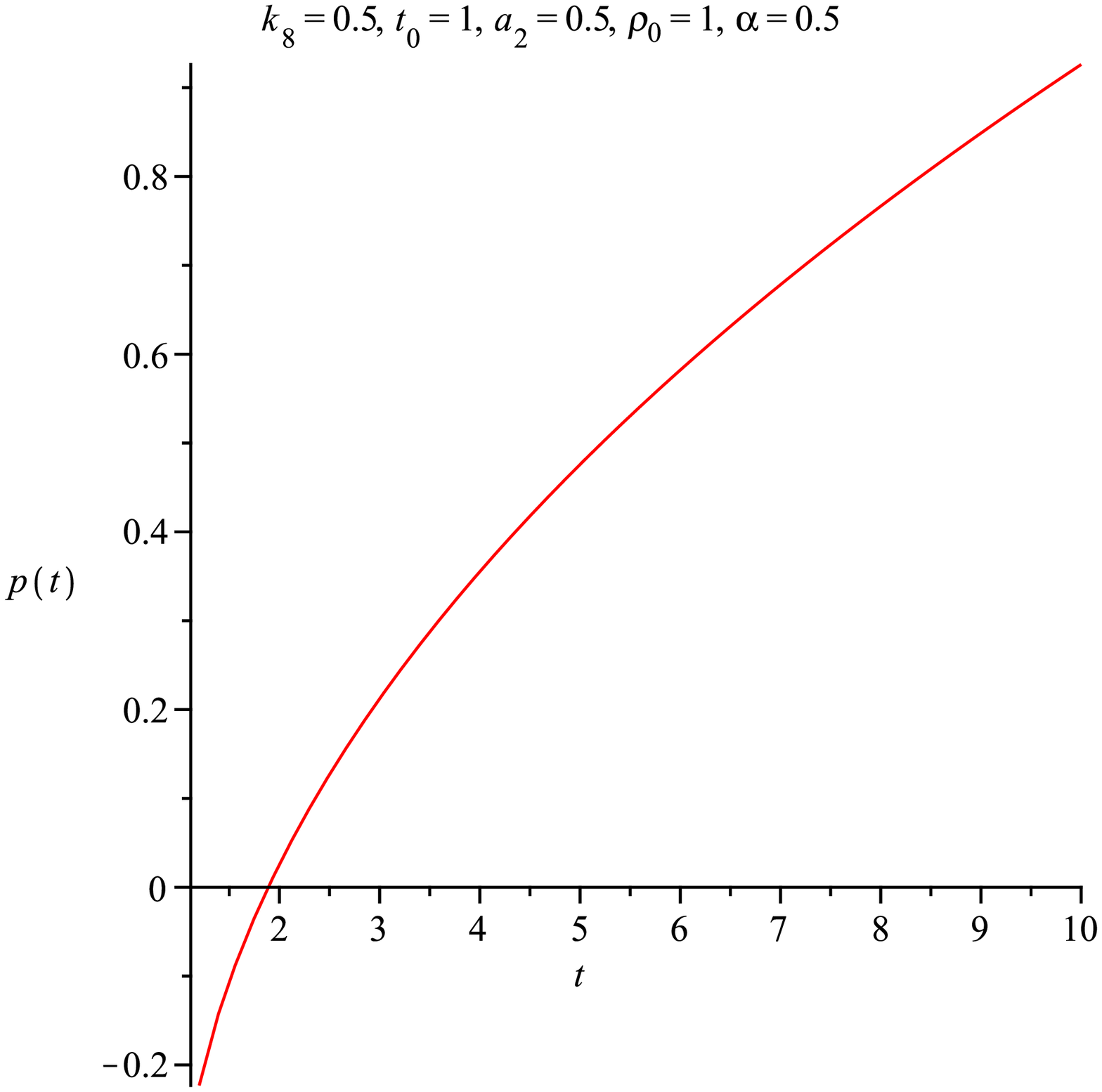}
\includegraphics[width=0.43\textwidth]{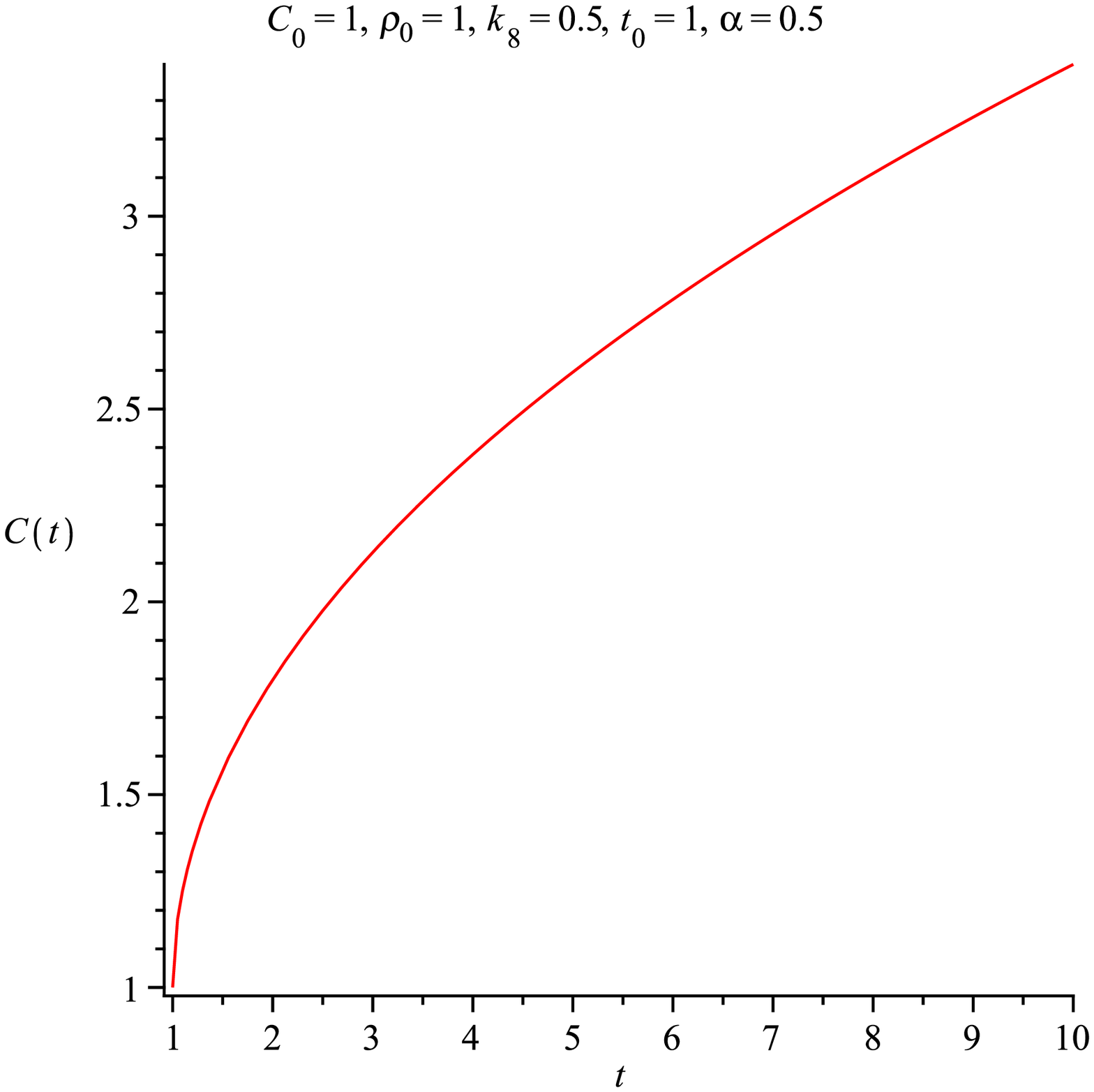}
\caption{Variation of pressure and creation field for Sub-case
4.2.3 when $\rho_0 \neq 0$, $a_2 \neq 0$, $\Omega_0
> 0$, $3\rho_0+9a_2^2=k_8^2$ and $\gamma=1$}
\end{center}
\end{figure}

\section{Non-singular solutions in the $C$-field cosmological models}

Here we assume
$\gamma=0,~\alpha^{2}=1,~a_{2}=0,~3\rho_{0}=-k_{0}{l}$, so that
\begin{equation}
V(t)= l+\frac{1}{4e^{2k_{1}{T}}} \left(e^{2k_{1}{T}} - l\right)^2,
\end{equation}
where $\frac{k_{0}}{4}={k_{1}}^2$ and $(t-t_{0})=T$. Also, we
obtain the following set of non-singular solutions:
\begin{equation}
\rho(t)=\frac{k_{0}}{12\pi}{\left[1-\frac{l}{2l+\frac{1}{2e^{2k_{1}{T}}}{\left(e^{2k_{1}{T}}-l\right)}^2}\right]}
\end{equation}

\begin{equation}
A(t)=a_1 \left[l+\frac{1}{4e^{2k_1 T}} \left(e^{2k_1 T} - l
\right)^2 \right]^{\frac{1}{3}}
\end{equation}

\begin{equation}
B(t)=\frac{1}{a_1^2} \left[l+\frac{1}{4e^{2k_1 T}} \left(e^{2k_1
T} - l \right)^2 \right]^{\frac{1}{3}}
\end{equation}

\begin{equation}
C(t)=C_{0}+\frac{1}{f\alpha}\sqrt{\frac{k_{0}}{12\pi}}{T}
\end{equation}

\begin{figure}
\begin{center}
\includegraphics[width=0.43\textwidth]{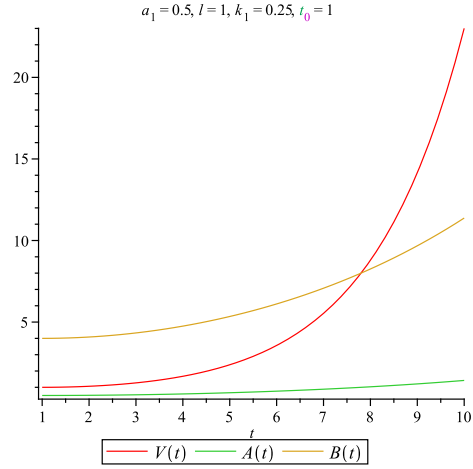}
\caption{Variation of volume $V$ and scale factors $A$ and $B$ for
non-singular case 5 when $\gamma = 0$, $a_2 = 0$, $\alpha^2 = 1$
and $3\rho_0 = - k_0 l$}
\end{center}
\end{figure}

\begin{figure}
\begin{center}
\includegraphics[width=0.43\textwidth]{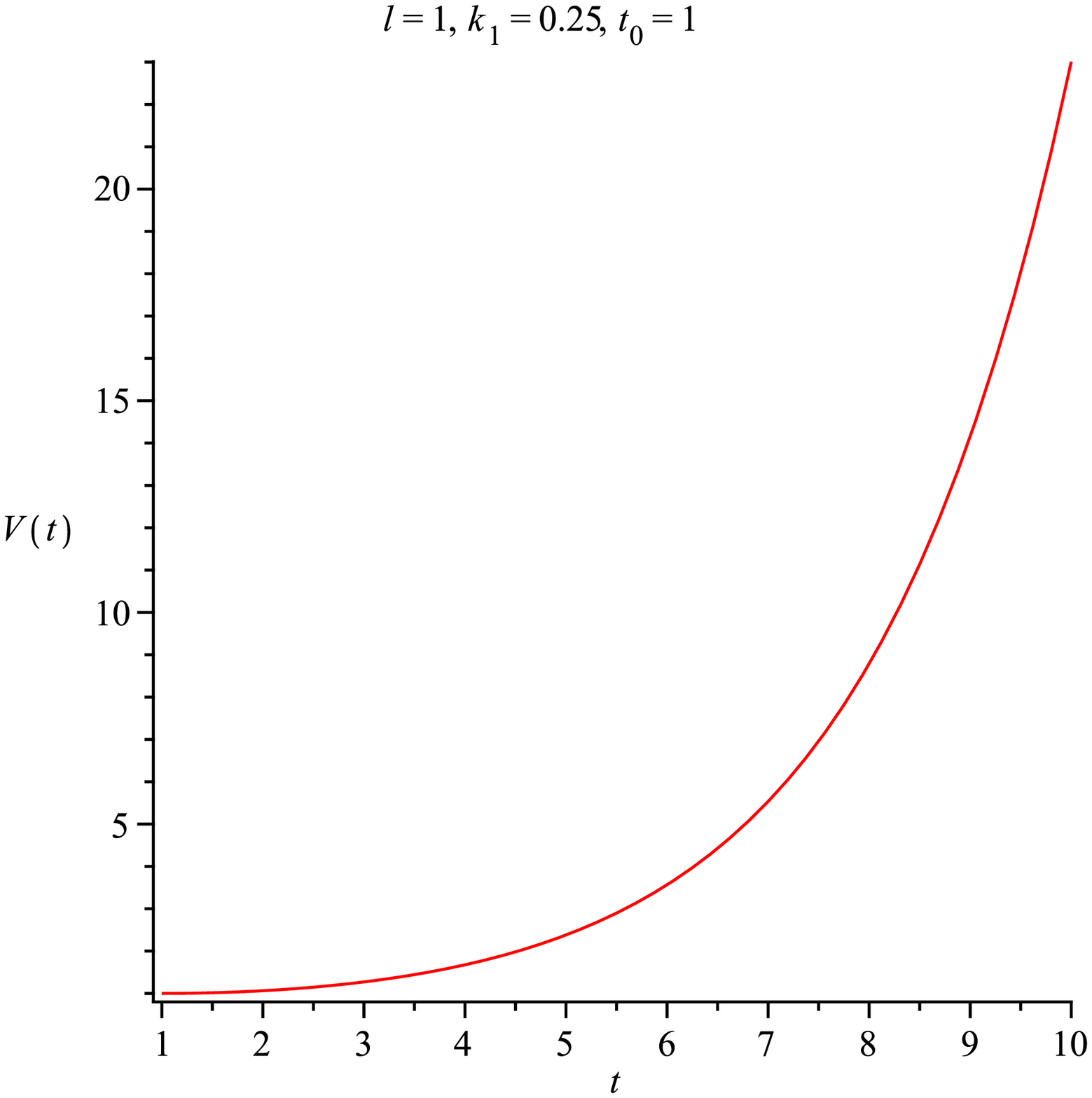}
\includegraphics[width=0.43\textwidth]{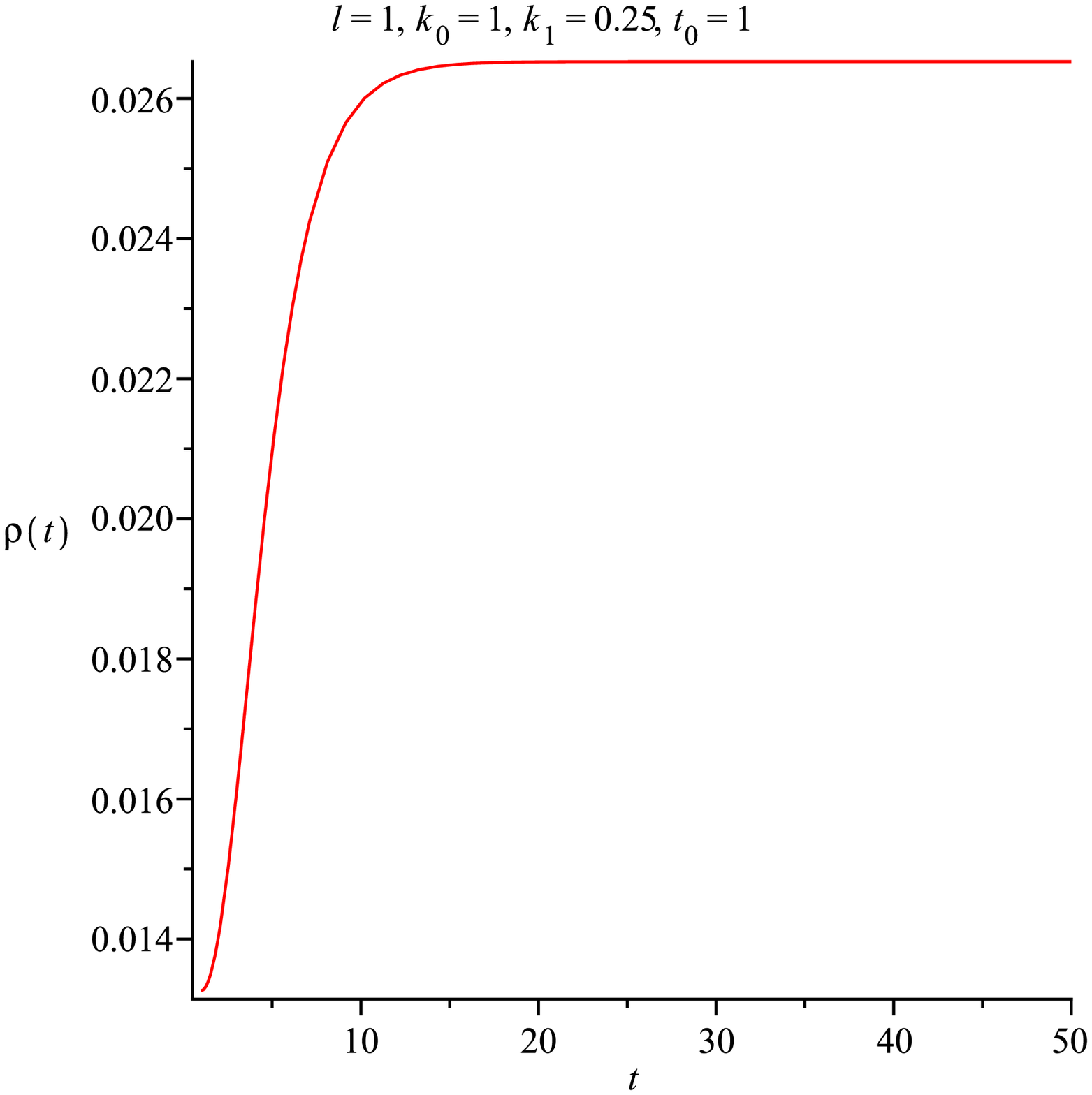}
\includegraphics[width=0.43\textwidth]{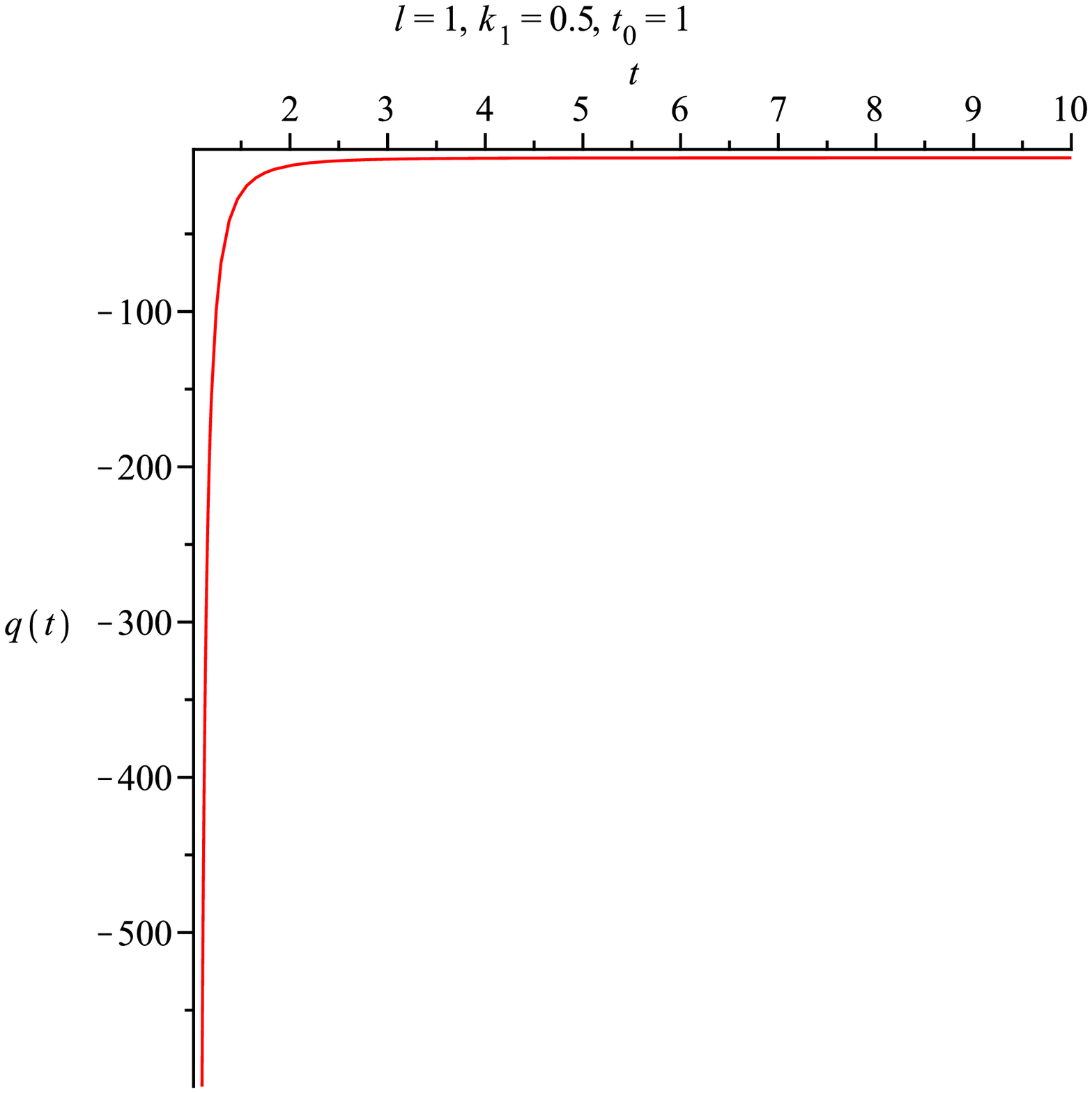}
\includegraphics[width=0.43\textwidth]{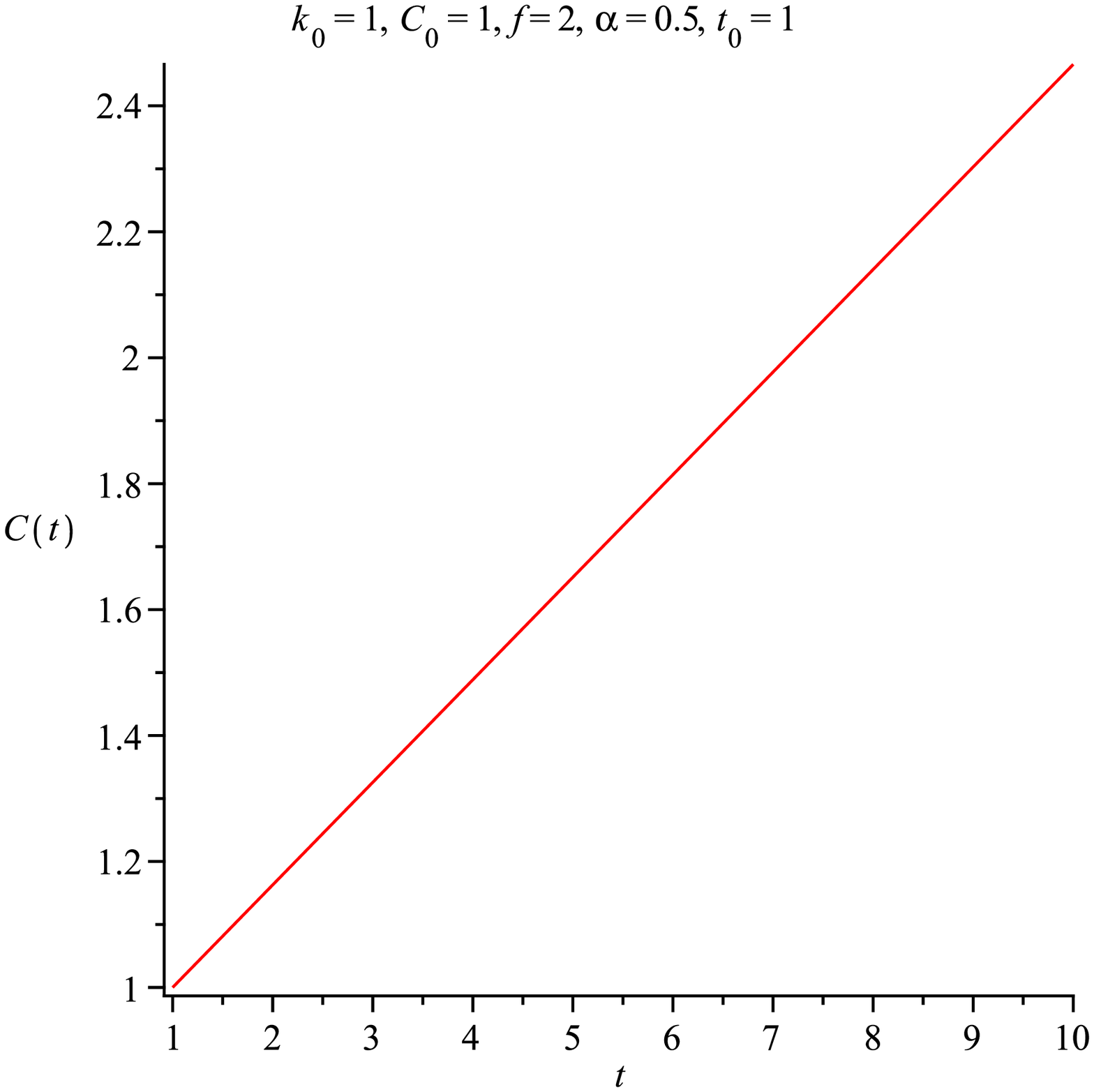}
\caption{Variation of volume, density, deceleration parameter and
creation field for non-singular case 5 when $\gamma = 0$, $a_2 =
0$, $\alpha^2 = 1$ and $3\rho_0 = - k_0 l$}
\end{center}
\end{figure}

\section{The physical properties of the models}

The expansion scalar is given by $\theta=3\,H$,
$H=\frac{\dot{a}}{a}=\frac{1}{3}\sum_{i=1}^{3}H_i$ is the Hubble
parameter in our anisotropic models, $a=V^{1/3}$ is the average
scale factor, $H_1=\frac{\dot{A}}{A}$, $H_2=\frac{\dot{B}}{B}$ and
$H_3=\frac{\dot{B}}{B}$ are the directional Hubble factors in the
directions of $x$, $y$ and $z$ respectively. The mean anisotropy
parameter is defined by
$\Delta=\frac{1}{3}\sum_{i=1}^{3}\Big(\frac{H_i}{H}-1\Big)^2$. The
shear scalar is given by
$\sigma^2=\frac{1}{2}\sum_{i=1}^{3}\Big(H_i^2-3H^2\Big)=\frac{3}{2}\,\Delta\,H^2$.e
The deceleration parameter is defined by $q=-(\frac{\dot{H}}{H^2}
+ 1)$.

It is evident that for all the cases discussed above, the shear
scalar is decreasing function of  time and finally diminishes for
sufficiently larger time except the sub-cases (4.1.1) \& (4.2.1).
For sub-cases (4.1.1) \& (4.2.1), the shear scalar is found to be
zero which is proposed for the model of non-shearing universe with
isotropic distribution.  However, the decreasing behaviour of
shear scalar corresponds to the isotropisation of universe with
passage of time. It is to note here that the direction Hubble's
Parameter measures the different rate of expansion along different
spatial directions at the same time which governs the anisotropy
of universe~(Kristian \& Sachs 1966, Collins et al. 1980, Saha \&
Yadav 2012, Yadav et al. 2012).

\subsection{The model {\it(4.1.1)}} In this case the
solution corresponds to:
\begin{equation}\label{u51}
  \begin{array}{ll}
\theta=\frac{1}{(1-\alpha^2)\,T},\,\,\,\,\,\,\,\Delta=\sigma^2=0,\,\,\,\,\,\,\,q=2-3\,\alpha^2.
  \end{array}
\end{equation}

\subsection{The model (i) of {\it(4.1.2)}} In this case the
solution corresponds to:
\begin{equation}\label{u52}
  \begin{array}{ll}
\theta=\frac{1}{T},\,\,\,\,\,\Delta=\frac{18\,a_2^2}{k_2^2},\,\,\,\,\,
\sigma^2=\frac{3\,a_2^2}{k_2^2\,T^2},\,\,\,\,\,q=2.
  \end{array}
\end{equation}

\subsection{The model (ii) of {\it(4.1.2)}} In this case the solution corresponds to:
\begin{equation}\label{u53}
  \begin{array}{ll}
\theta=\frac{2\,T}{T^2-k_3^2},\,\,\,\,\,\Delta=\frac{2\,k_3^2}{T^2},\,\,\,\,\,
\sigma^2=\frac{4\,k_3^2}{3\,\Big(T^2-k_3^2\Big)^2},\,\,\,\,\,q=\frac{1}{2}(1+\frac{3\,k_3^2}{T^2}).
  \end{array}
\end{equation}

\subsection{The model (iii) of {\it(4.1.2)}} In this case the solution
corresponds to:

\begin{equation}\label{u54}
  \begin{array}{ll}
\theta=a \,a_2\,k_4\,
\coth\big[3\,a_2\,k_4\,T\big],\,\,\,\,\,\,\,\,\,\,
\Delta=2\,\mathrm{sech}^2\big[3\,a_2\,k_4\,T\big],\\
\\
\sigma^2=2\,a_2^2\,k_4^4\,\mathrm{csch}^2\big[3\,a_2\,k_4\,T\big],\,\,\,\,\,\,\,\,\,\,
q=3\,\mathrm{sech}^2\big[3\,a_2\,k_4\,T\big]-1.
  \end{array}
\end{equation}

\subsection{The model {\it(4.2.1)}} In this case the solution
corresponds to:

\begin{equation}\label{u55}
  \begin{array}{ll}
\theta=\frac{2}{(1+\gamma)\,T},\,\,\,\,\,\,\,\Delta=\sigma^2=0,\,\,\,\,\,\,\,q=\frac{1+3\,\gamma}{2}.
  \end{array}
\end{equation}

\subsection{The model (i) of {\it(4.2.2)}} In this case the solution
corresponds to:
\begin{equation}\label{u56}
  \begin{array}{ll}
\theta=\frac{2\,T}{T^2-k_6^2},\,\,\,\,\,\Delta=\frac{8\,a_2^2}{\rho_0^2\,T^2},\,\,\,\,\,
\sigma^2=\frac{16\,a_2^2}{3\,\rho_0^2\,\Big(T^2-k_6^2\Big)^2},\,\,\,\,\,q=\frac{1}{2}(1+\frac{3\,k_6^2}{T^2}).
  \end{array}
\end{equation}

\subsection{The model (ii) of {\it(4.2.2)}} In this case the solution
corresponds to:
\begin{equation}\label{u57}
  \begin{array}{ll}
\theta=k_7\Bigg(\frac{\mathrm{e}^{2\,k_7\,T}-a_3}{\mathrm{e}^{2\,k_7\,T}-6\,\rho_0\,\mathrm{e}^{k_7\,T}+a_3}\Bigg),\,\,\,\,\,\,\,\,\,\,
\Delta=\frac{8\,(9\,\rho_0^2-a_3)\,\mathrm{e}^{2\,k_7\,T}}{\Big(\mathrm{e}^{2\,k_7\,T}-a_3\Big)^2},\\
\\
\sigma^2=\frac{4\,k_7^2\,(9\,\rho_0^2-a_3)\,\mathrm{e}^{2\,k_7\,T}}{3\,\Big(\mathrm{e}^{2\,k_7\,T}-6\,\rho_0\,\mathrm{e}^{k_7\,T}+a_3\Big)^2},\\
\\
q=-\frac{\mathrm{e}^{4\,k_7\,T}-18\,\rho_0\,\mathrm{e}^{3\,k_7\,T}+2\,a_3\,\Big(5\,\mathrm{e}^{2\,k_7\,T}
-9\,\rho_0\,\mathrm{e}^{k_7\,T}\Big)+a_3^2}{\Big(\mathrm{e}^{2\,k_7\,T}-a_3\Big)^2}.
  \end{array}
\end{equation}
where $a_3=9\,\rho_0^2-36\,a_2^2\,k_7^2$.

\subsection{The model {\it(4.2.3)}} In this case the solution
corresponds to:
\begin{equation}\label{u58}
  \begin{array}{ll}
\theta=\frac{1}{T},\,\,\,\,\,\Delta=\frac{18\,a_2^2}{k_8^2},\,\,\,\,\,
\sigma^2=\frac{3\,a_2^2}{k_8^2\,T^2},\,\,\,\,\,q=2.
  \end{array}
\end{equation}

\subsection{The model {\it(5)}} In this case the solution
corresponds to:
\begin{equation}\label{u59}
  \begin{array}{ll}
\theta=\frac{k_1\left(e^{2k_1T}-l\right)}{l+\frac{1}{4e^{2k_1 T}}
\left(e^{2k_1 T} - l \right)^2}
,\,\,\,\,\,\Delta=\frac{\left(e^{2k_1T}-l\right)^2}{4e^{4k_{1}T}}
,\,\,\,\,\,\sigma^2=\frac{k_{1}^2\left(e^{2k_1T}-l\right)^4}{24e^{4k_1T}\left[l+\frac{1}{4e^{2k_1
T}} \left(e^{2k_1 T} - l \right)^2 \right]^2}
,\,\,\,\,\,

\\q=\frac{3l\left(l^2 e^{-2k_1T} -
3e^{2k_1T}-2l\right)}{2\left(e^{2k_1T} -l \right)^2}-1.
 \end{array}
\end{equation}

\section{Discussions and Conclusions}
In the present work plane symmetric space-time filled with perfect
fluid in the Hoyle-Narlikar $C$-field cosmology has been
investigated. By considering (i) the creation field is a function
of time alone, and (ii) the rate of creation of matter
energy-density is proportional to the strength of the existing
$C$-field energy-density we have found out a new class of exact
solutions.

We have, in general, discussed several physical features and
geometrical properties of the models. However, as a special case,
most notable aspect of the solution set that have been studied are
non-singular in nature. These aspects have been shown through
several plots which are of two kinds: Figs. 1 - 10 for singular
cases and Figs. 11 - 12 for non-singular case. All figures depict
interesting features of the present cosmological model in terms of
$C$-field and other physical parameters.

\begin{figure}
\begin{center}
\includegraphics[width=0.43\textwidth]{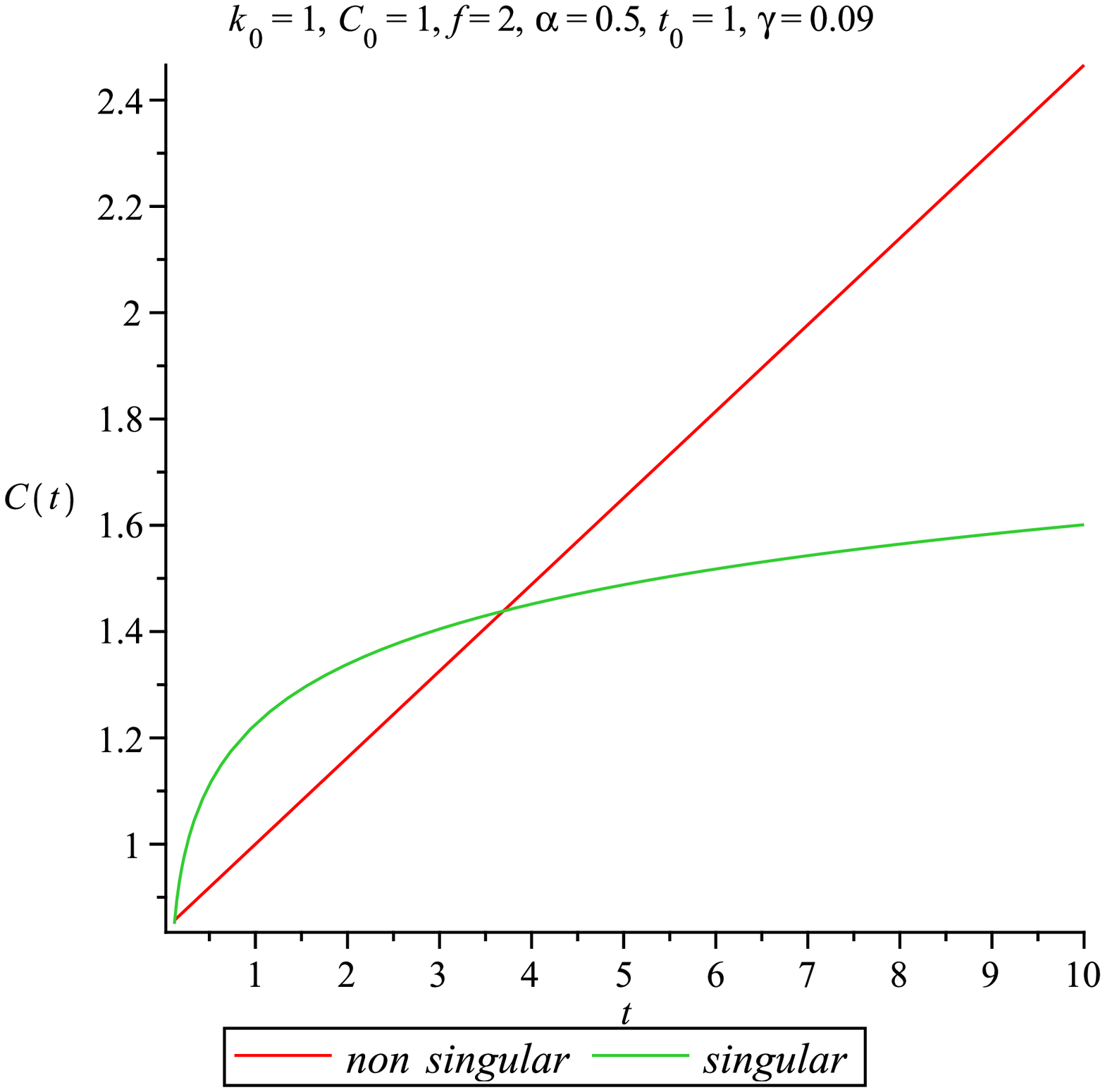}
\includegraphics[width=0.43\textwidth]{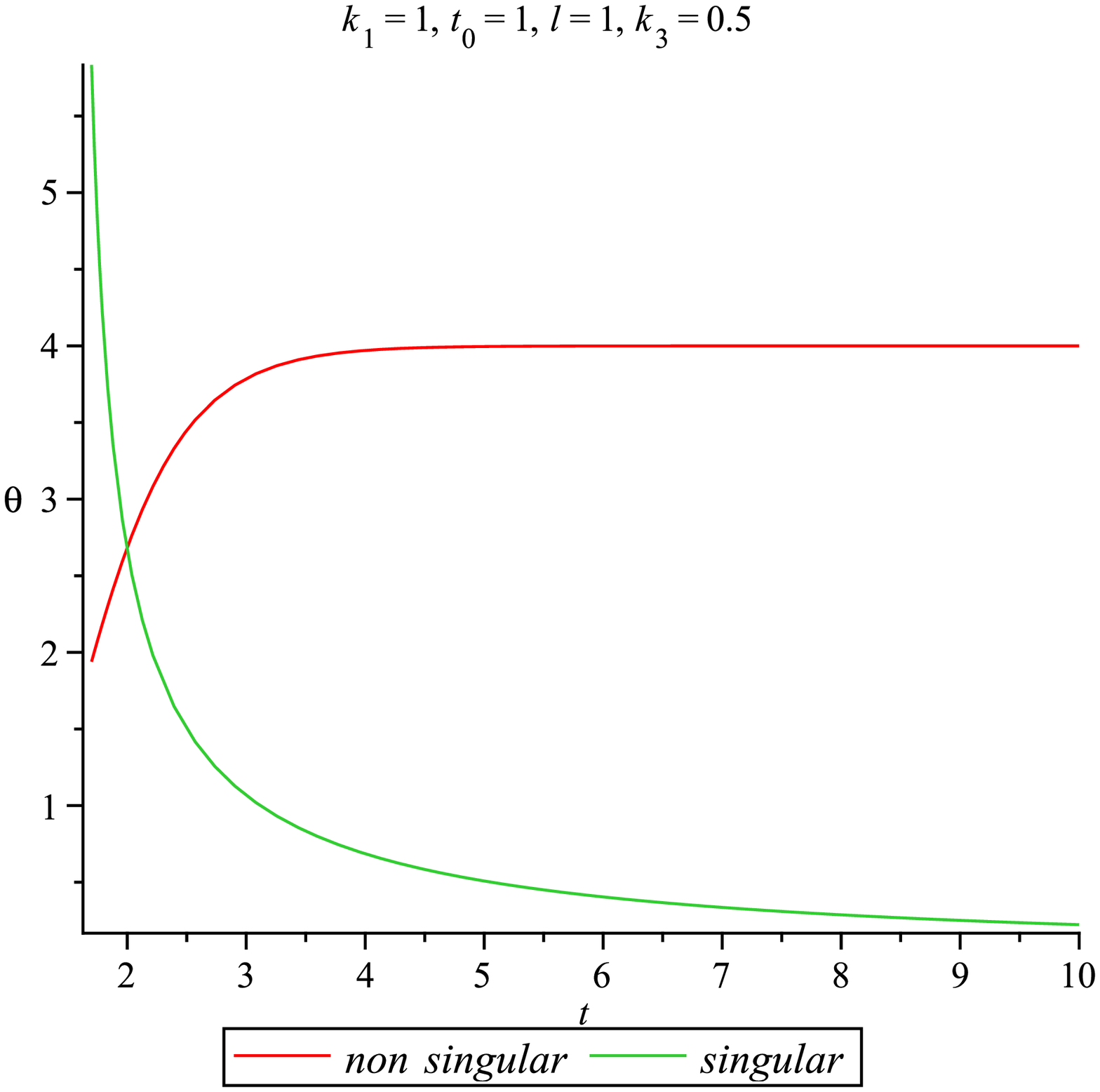}
\includegraphics[width=0.43\textwidth]{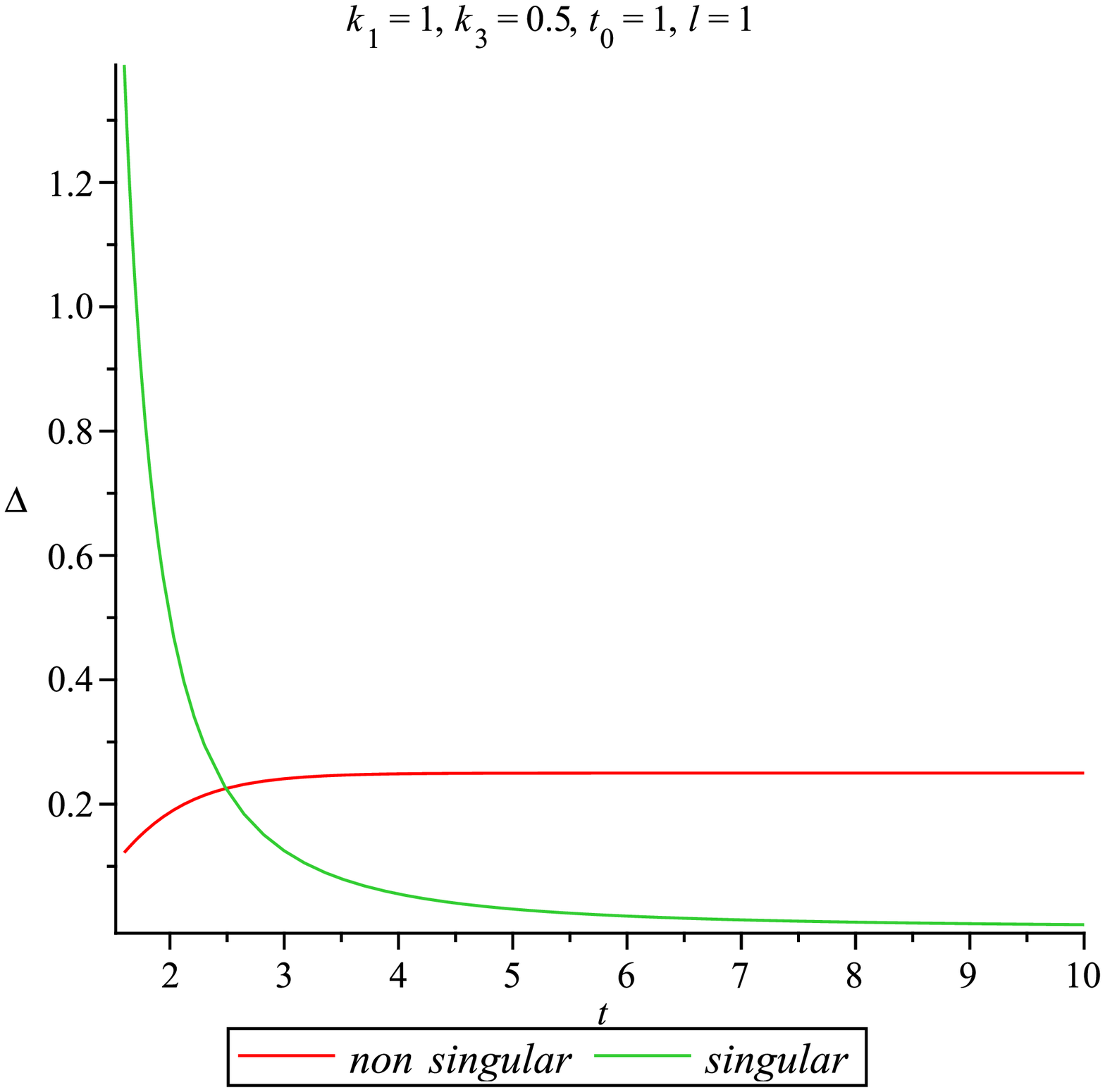}
\includegraphics[width=0.43\textwidth]{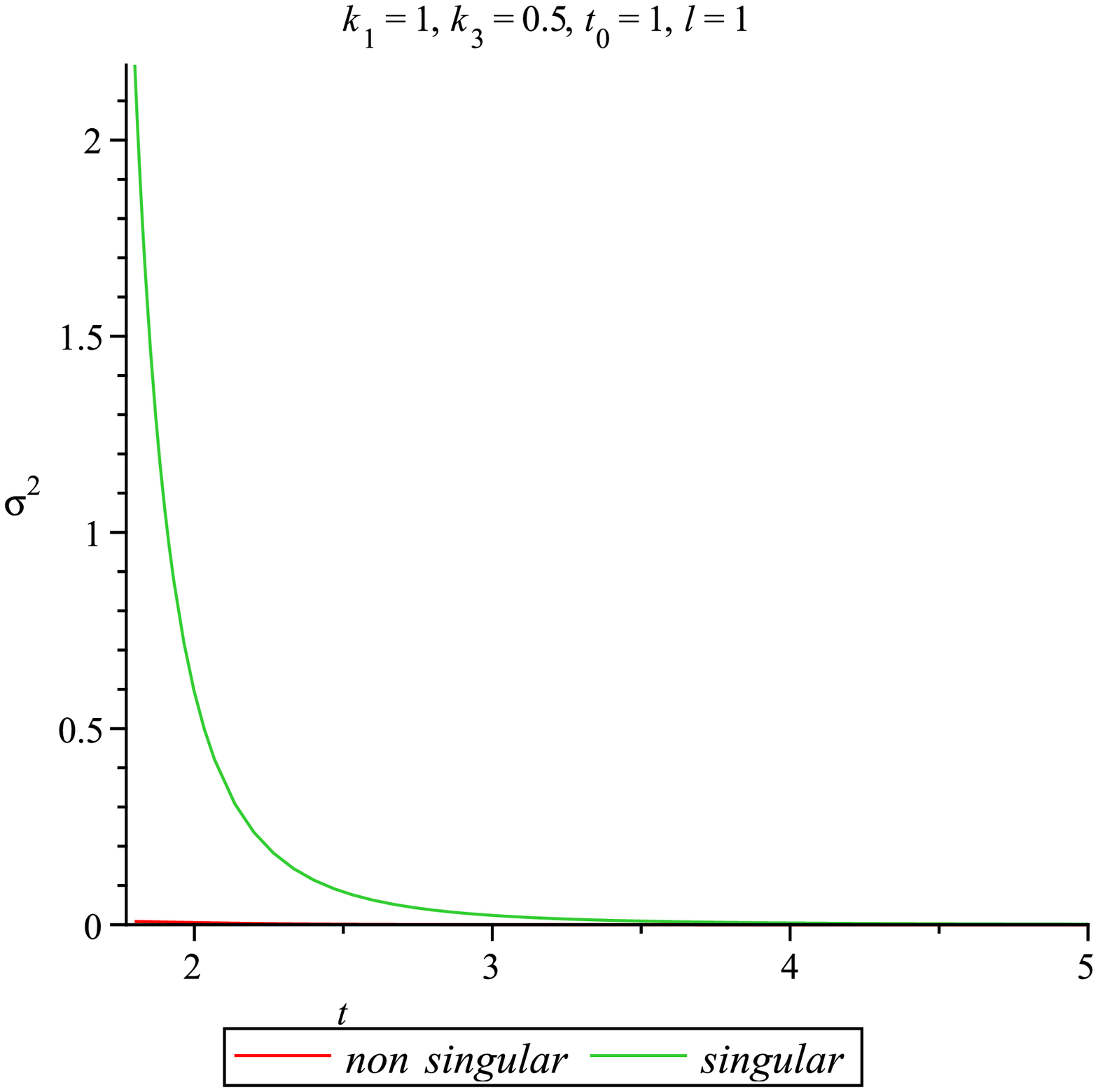}
\includegraphics[width=0.43\textwidth]{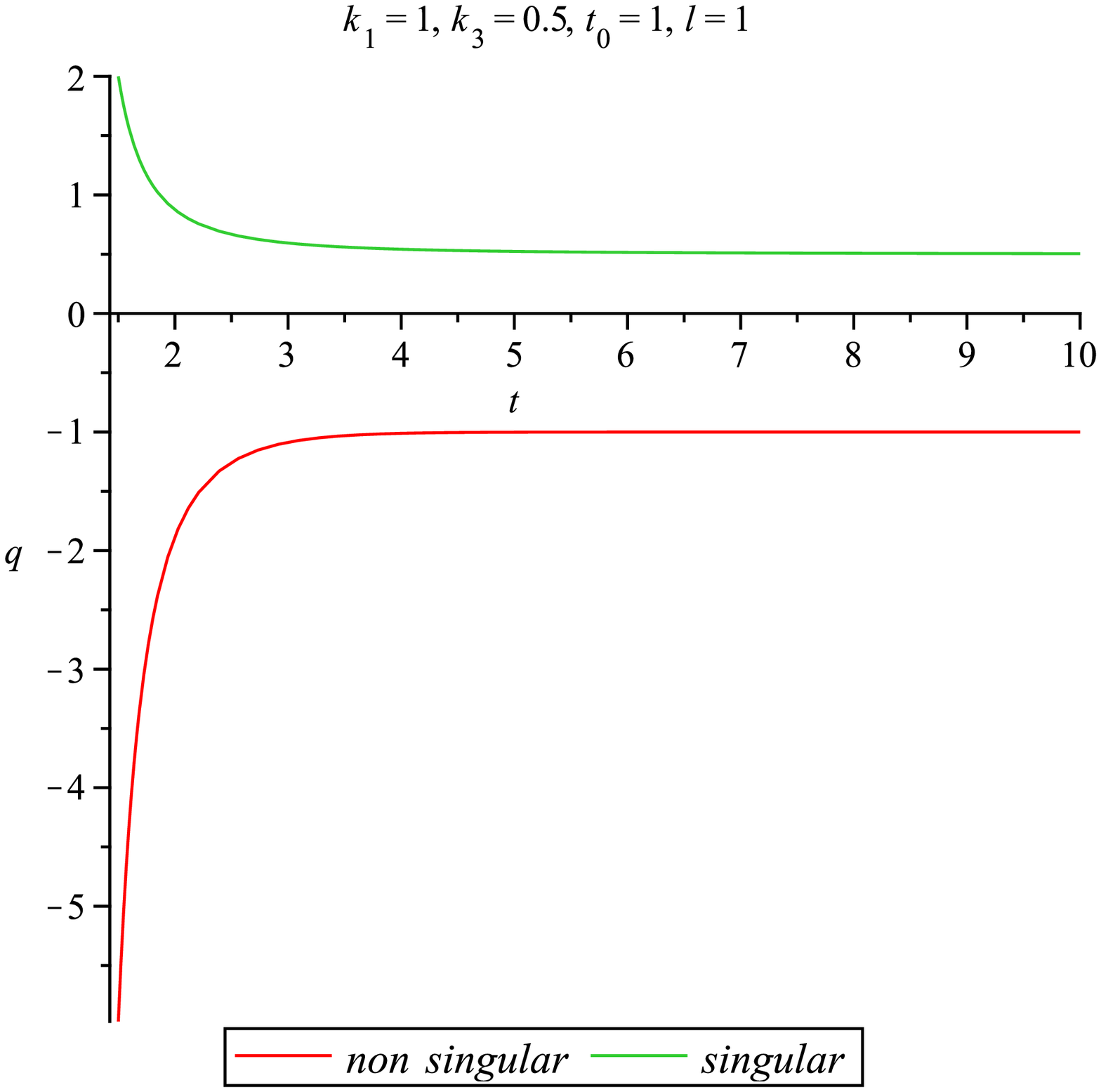}
\caption{Variation of $C$, $\theta$, $\Delta$, $\sigma^2$ and $q$
for singular and non-singular cases as a comparative study}
\end{center}
\end{figure}

However, as one possible improvement of the present investigation
we would like to perform a comparative study between the singular
and non-singular solutions of the two models. In this regard we
draw a few specific plots to show variation of $C$, $\theta$,
$\Delta$, $\sigma^2$ and $q$ for singular and non-singular cases
in Figs. 13. Here we are basically doing comparison of the
singular case 4.1.2 (ii) with non-singular case 5. One can observe
that in the model of singular case the decelerating parameter $q$
gets positive value whereas non-singular model gives an
accelerating Universe. Comparing $C$ in both the cases we note
that initially the creation field in singular case assumes higher
value than non-singular one, however, after lapsing certain time
the creation field in non-singular case acquires higher value than
singular one. In a similar way one can continue comparison for
other parameters also which is quite evident from the contrasting
behaviour of other parameters in Fig. 13.

As a final comment, we note from the above comparative study that
the present model in a unique way represents both the features of
the accelerating as well as decelerating Universe depending on the
parameters and thus seems provides glimpses of the oscillating or
cyclic model of the Universe (see Frampton~\cite{Frampton2006} and Refs.
therein). However, it is to be noted that our model is based on
Hoyle-Narlikar type $C$-field cosmological theory and does not
invoke any other agent or theory, e.g. dark
energy (Khoury et al.~\cite{Khoury2001}; Steinhardt \& Turok~\cite{Steinhardt2002a};
Steinhardt \& Turok~\cite{Steinhardt2002b}; Boyle~\cite{Boyle2004}; Steinhardt \& Turok~\cite{Steinhardt2006}),
branes (Randall \& Sundrum~\cite{Randall1999a,Randall1999b}; Csaki et al.~\cite{Csaki2000};
Binetruy et al.~\cite{Binetruy2000}; Brown et al.~\cite{Brown2004}),
modified gravity (Frampton \& Takahashi~\cite{Frampton2003,Frampton2004}) etc. in allowing
cyclicity.

\begin{acknowledgements}
We all are extremely grateful to Prof. J.V. Narlikar for his
valuable suggestions and constant inspiration which have enabled
us to improve the manuscript substantially. SR and FR are thankful
to the authority of Inter-University Centre for Astronomy and
Astrophysics, Pune, India for providing the Associateship
programme under which a part of this work was carried out. IHS is
also thankful to DST, Government of India, for providing financial
support under INSPIRE Fellowship. We are very grateful to an
anonymous referee for his/her insightful comments that have led to
significant improvements, particularly on the aspects of
interpretation.
\end{acknowledgements}

\end{document}